\documentclass[twocolumn]{aastex631}

\usepackage[version=4]{mhchem}

\begin{document}

\title{Water Cloud and Chemical Modulations in the Coldest Brown Dwarf}

\author{Brittany E.~Miles}\affiliation{Steward Observatory, University of Arizona, 933 N. Cherry Ave., Tucson, AZ 85721, USA}
\author{James Mang  }\altaffiliation{NSF Graduate Research Fellow} \affiliation{Department of Astronomy, University of Texas at Austin, Austin, TX 78712, USA}
\author{Caroline V. Morley  }\affiliation{Department of Astronomy, University of Texas at Austin, Austin, TX 78712, USA}
\author{Andrew J. I. Skemer}\affiliation{Department of Astronomy and Astrophysics, University of California, Santa Cruz, 1156 High St, Santa Cruz, CA 95064, USA}
\author{Melanie Rowland}\affiliation{Department of Astrophysics, American Museum of Natural History, Central Park West at 79th Street, New York, NY 10024, USA}
\author{Brianna Lacy}\affiliation{NASA Ames Research Center Space Science and Astrobiology Division Moffett Field, CA 94035, US}

\author{Genaro Suarez}\affiliation{ Department of Astrophysics, American Museum of Natural History, Central Park West at 79th Street, New York, NY 10024, USA}
\author{Channon Visscher}\affiliation{ Chemistry \& Planetary Sciences, Dordt University, Sioux Center, IA, USA} \affiliation{Center for Extrasolar Planetary Systems, Space Science Institute, Boulder, CO, USA}

\author{Dániel Apai}\affiliation{Steward Observatory, University of Arizona, 933 N. Cherry Ave., Tucson, AZ 85721, USA} \affiliation{ Lunar and Planetary Laboratory, University of Arizona, 1629 East University Boulevard, Tucson, AZ 85721, USA}

\author{Gordon Bjoraker}\affiliation{University of Maryland Baltimore County, Baltimore MD 21250, USA}

\author{Aarynn L. Carter}\affiliation{Space Telescope Science Institute (STScI), 3700 San Martin Drive, Baltimore, MD 21218, USA}

\author{Jacqueline K. Faherty}\affiliation{ Department of Astrophysics, American Museum of Natural History, Central Park West at 79th Street, New York, NY 10024, USA}

\author{Jonathan J. Fortney}\affiliation{Department of Astronomy and Astrophysics, University of California, Santa Cruz, 1156 High St, Santa Cruz, CA 95064, USA}

\author{Nguyen Fuda}\affiliation{Lunar and Planetary Laboratory, University of Arizona, 1629 E. University Boulevard, Tucson, AZ 85721, USA}

\author{Thomas Geballe} \affiliation{Gemini Observatory/NSF’s NOIRLab, 670 N. Aohoku Place, Hilo, HI, 96720, USA}

\author{Harshil Kothari}\affiliation{Ritter Astrophysical Research Center, Department of Physics \& Astronomy, University of Toledo, 2801 W. Bancroft Street, Toledo, OH 43606, USA}

\author{Mary Anne Limbach} \affiliation{Department of Astronomy, University of Michigan, Ann Arbor, MI 48109, USA}

\author{Mark Marley}\affiliation{ Lunar and Planetary Laboratory, University of Arizona, 1629 East University Boulevard, Tucson, AZ 85721, USA}

\author{Allison M. McCarthy}\affiliation{School of Physics, Trinity College Dublin, The University of Dublin, Dublin 2, Ireland}

\author{Adam Schneider}\affiliation{United States Naval Observatory, Flagstaff Station, 10391, West Naval Observatory Road, Flagstaff, AZ 86005, USA}

\author{Johanna Vos}\affiliation{School of Physics, Trinity College Dublin, The University of Dublin, Dublin 2, Ireland}

\author{Michael Cushing}\affiliation{Ritter Astrophysical Research Center, Department of Physics and Astronomy, University of Toledo, Toledo, OH 43606, USA}
\author{Richard Freedman}\affiliation{SETI Institute, Mountain View, CA 94035, USA}
\author{Michael Line}\affiliation{School of Earth and Space Exploration, Arizona State University, Tempe, AZ 85281, USA}
\author{Roxana Lupu}\affiliation{Eureka Scientific, Inc, Oakland, CA 94602}
\author{Emily Martin}\affiliation{Department of Astronomy and Astrophysics, University of California, Santa Cruz, 1156 High St, Santa Cruz, CA 95064, USA}

\author{Mikayla J. Wilson}\altaffiliation{NSF Graduate Research Fellow} \affiliation{Department of Astronomy and Astrophysics, University of California, Santa Cruz, 1156 High St, Santa Cruz, CA 95064, USA}

\begin{abstract}
We present high signal-to-noise (80 -- 100), medium resolution (R $\sim$1000), time-series JWST/NIRSpec spectra of WISE 0855 (265K), the coldest known brown dwarf. Medium resolution, time-series spectroscopy gives us the power to disentangle the effects of chemistry, temperature, and condensates on this cool world. Our observations span 11 hours with a 15 minute cadence covering 2.87–5.27 $\mu$m. The strongest time variable spectroscopic feature is carbon monoxide gas absorption producing modulations with a peak-to-peak amplitude up to 10\% at some wavelengths. Using principal component analysis, we show that the variations in carbon monoxide and phosphine correlate with one another. By comparing our data to atmospheric and structure models we present evidence of patchy water clouds within the atmosphere of WISE 0855. We find that variations in \ce{CO and PH3} abundances must originate from quenched atmospheric pressures while variations in water cloud thickness occurs at lower pressures.
\end{abstract}

\keywords{Brown Dwarfs, Exoplanets, Gas-Giant Exoplanets}

\section{Introduction}
Astronomers have been studying brown dwarfs with the goal of understanding how atmospheric dynamics impact the 3-dimensional structure of clouds and chemistry within their atmospheres for over two decades. Inhomogeneous atmospheric features rotate in and out of view, causing the integrated emission of a brown dwarf to change in brightness over time. Pre-JWST time-series studies of brown dwarfs have been limited to broadband photometry or relatively low resolution spectroscopy \citep{1999MNRAS.304..119T,2001A&A...367..218B,2001A&A...374.1071B, 2002ApJ...577..433G,2012ApJ...750..105R, 2013ApJ...768..121A,2016ApJ...832...58E,2019MNRAS.483..480V,2020AJ....160...38V,2022ApJ...924...68V, 2021MNRAS.503..743B, 2024MNRAS.527.6624L, 2024ApJ...970...62P, 2026arXiv260626411C}. In the JWST era time-resolved, medium-resolution spectroscopy beyond the near-infrared is finally possible for the coldest ($T_{\rm eff}\lesssim$ 450 K) brown dwarfs known as Y dwarfs. 

Y dwarfs, which were discovered by the Wide-field Infrared Survey Explorer (WISE) mission \citep{2010AJ....139.2455E, 2011ApJ...740..108L, 2011ApJ...743...50C}, have hydrogen dominated atmospheres with significant absorption features from methane (\ce{CH4}) and ammonia (\ce{NH3}) \citep{2012ApJ...753..156K, 2023ApJ...951L..48B, 2024ApJ...973..107B, 2024ApJ...973...60B, 2024AJ....167..237L}. Vertical convection transports relatively warm gas into the upper atmosphere, enriching the apparent abundances of carbon dioxide (\ce{CO2}) and carbon monoxide (\ce{CO}) in nearly all Y dwarfs \citep{2020AJ....160...63M,2023ApJ...951L..48B, 2024ApJ...973..107B, 2024ApJ...973...60B, 2024AJ....167..237L}. Some Y-dwarfs also show enriched abundances of phosphine (\ce{PH3}) \citep{2024ApJ...977L..49R,doi:10.1126/science.adu0401}. The coolest Y-dwarfs ($\lesssim$375 K) are predicted to have water condensation at the very top of their upper atmospheres \citep{1997ApJ...491..856B,2003ApJ...596..587B, 2014ApJ...787...78M, 2023ApJ...950....8L}. WISE J085510.83-071442.5 (hereinafter referred to as WISE 0855, 265 K) \citep{2014ApJ...786L..18L} is cool enough to host water clouds in its upper atmosphere but warm enough to prevent the condensation of \ce{NH3} \citep{1997ApJ...491..856B}, making this object one of the best laboratories for disentangling the impact of longitudinal water cloud structure and vertical mixing outside of our Solar System.

In this paper, we present time-series spectroscopic observations of WISE 0855 using JWST/NIRSpec. In Section~\ref{sec:WISE0855} we cover the previous work and interpretations of WISE 0855's atmosphere. We describe the observational strategy and data reduction of our program in Section~\ref{sec:Observations and Data Reduction}. We examine the light curves and qualitatively identify the main molecular features that change over time in Section~\ref{sec:Features}. In Section~\ref{sec:Analysis} we analyze the JWST time-series spectra with principal component analysis and use two Spitzer epochs and our JWST observations to estimate the periodicity of WISE 0855. Atmospheric models are applied to the average and time-series data in Section~\ref{sec:Model Fitting}. Finally, in Section~\ref{sec:Discussion} we discuss the implications of our results and the potential surface features that can explain WISE 0855's brightness variations.  

\section{WISE 0855} \label{sec:WISE0855}

The presence and composition of clouds on WISE 0855 have been discussed extensively in the literature. Previous work interpreting the atmosphere of WISE 0855 has found that either clouds or disequilibrium chemistry could broadly, but imperfectly, match available data. The earliest evidence for clouds was presented with the first near-infrared detection of WISE 0855 in \cite{2014ApJ...793L..16F} where the absolute WISE W2 photometry and J-band -- W2 colors were found to be consistent with atmospheric models from \cite{2012ApJ...756..172M} which include water and sulfide clouds. Following this work, \cite{2014ApJ...796....6L} showed WISE 0855's photometry can also be consistent with altered clear, disequilibrium models from \cite{Saumon_2012}. 

Further work by \cite{2016ApJ...823L..35S} and \cite{2016AJ....152...78L} presented optical through mid-infrared photometry of WISE 0855. Their results showed that no model grid is able to reproduce the photometric colors of WISE 0855 or the entire Y-dwarf sequence.

The first spectroscopic observations of WISE~0855, presented in \cite{2016ApJ...826L..17S}, were capable of disentangling the impact of disequilibrium chemistry and clouds that was previously intertwined within the W2 bandpass. In \cite{2016ApJ...826L..17S} the addition of a gray opacity source where water clouds are expected was a better fit to WISE 0855's low-resolution M-band spectra compared to the cloudless atmospheric model. \cite{2016ApJ...832...58E} published the first time-series photometric study of WISE 0855 using Spitzer's IRAC~1 and IRAC~2 channels. They concluded that hot spots are unlikely and only a few percent change in cloud coverage could explain the observed amplitude variability in both Spitzer channels, but maintain that better models are needed to interpret the underlying physics of the variability. \cite{2018ApJ...858...97M} introduced WISE 0855's L-band spectrum and compared the optical through mid-infrared data with self-consistent models that include water clouds. This work showed that water clouds could explain the spectrum's deviations from a clear atmosphere at wavelengths longer than 1.6 $\mu$m and a deep, gray opacity source at a pressure near 11 bars can explain the discrepancy within optical/near-infrared photometric bands.

\cite{2024AJ....167....5L} was an initial JWST-era study of WISE 0855 that compared atmospheric models to the object's 1 -- 5 $\mu$m spectrum. Their preliminary modeling determined that a disequilibrium, cloudless model with an adjusted adiabatic index can match WISE 0855's major spectral features but with discrepancies between 2 -- 3~$\mu$m.  Following this study, \cite{2024ApJ...977L..49R} completed two separate cloudless atmospheric retrievals on the medium resolution JWST/NIRSpec spectrum presented in \cite{2024AJ....167....5L} and the time averaged 3 -- 5~$\mu$m spectrum from the data we will present in this paper. The retrievals could only fit both independent data sets with a flexible pressure-temperature profile approach. The work in \cite{2024ApJ...977L..49R} showed the 4 -- 5 $\mu$m region, which has previously been cited as evidence for water clouds can be modeled only using disequilibrium molecular abundances however, the pressure-temperature profile deviates significantly from radiative-convective equilibrium, indicating a missing opacity source that could be attributed to clouds. The analysis in \cite{2024ApJ...977L..49R} also constrained the abundances of phosphine (\ce{PH3}) and deuterated methane (\ce{CH3D}) in a brown dwarf atmosphere for the first time ever.

The broadest wavelength characterization of WISE 0855 to-date is published in \cite{2025A&A...695A.224K}, which presented the first spectrum of the object from 5 -- 22 $\mu$m. Using both cloudy and clear atmospheric retrievals, they conclude WISE 0855's upper atmosphere water abundance is depleted above a pressure of 3 bars, but mention water vapor depletion does not directly imply water condensation. Their best fit cloudy retrieval suggests a cloud at 10 bars, which is much deeper than 1.2 -- 2.0 bars where water is expected to condense \citep{2014ApJ...787...78M, 2023ApJ...950....8L}.  Similarly to \cite{2024ApJ...977L..49R} the best fit pressure-temperature profiles deviate significantly from adiabatic and the authors suggest using a more flexible pressure-temperature profile framework in future work. The best fit object radii from the retrievals in \cite{2025A&A...695A.224K},  \cite{2024ApJ...977L..49R}, and \cite{2024AJ....167....5L} range between 0.79 -- 0.89 R$_{\rm jup}$ and are too small compared to the 1 R$_{\rm jup}$ expected based on evolutionary models for an object of WISE 0855's temperature with an assumed age between 1 and 10 Gyr \citep{2008ApJ...689.1327S}. \cite{kuhnle_2025_14762536} also compares a variety of  self-consistent forward models cloudy, clear, and disequilibrium models against one another. No single atmospheric, forward model can fit the entire 1 -- 22 $\mu$m spectrum of WISE 0855 however, the cloudy, disequilibrium models from \cite{2023ApJ...950....8L} provides the best overall fit to the 1 -- 22 $\mu$m spectrum. 

Since WISE 0855 has already been observed to be variable using broadband Spitzer photometry, this implies that one or more atmospheric parameters are inhomogeneous across the surface of the object. Compared to previous time-averaged spectra, our time-series JWST/NIRSpec spectral observations resolve variability across individual molecular features, allowing us to distinguish feature-specific variability from broadband changes that may arise from cloud or temperature modulations on the surface \citep{2014ApJ...787...78M, 2018ApJ...858...97M, 2022ApJ...927..184M}. This data set offers a new perspective on whether or not WISE 0855 has clouds and provides evidence for the physical origin of its rotational modulations.

\section{Observations and Data Reduction} \label{sec:Observations and Data Reduction}
time-series spectroscopic observations of WISE 0855 were taken using NIRSpec on JWST in Bright Object time-series (BOTS) mode for the Cycle 1 JWST GO program 2327 \citep{2021jwst.prop.2327S}. This mode uses the S1600A1 slit with a 1.6” × 1.6” square aperture to maximize throughput and minimize losses from drift. No dithering is used to preserve photometric stability. WISE 0855 was acquired using the NIRSpec Wide Aperture Target Acquisition (WATA) sequence then followed up with a confirmation and adjustment image in the S1600A1 slit. The time-series observations used the G395M/F290LP grating/filter combination to cover 2.87–5.27 $\mu$m at a resolution of $\sim$1000. The detector was set up using the NRSRAPID readout with 996 groups per integration, 44 integrations per exposure resulting in an 11 hour sequence with 15-minute cadence. The spectroscopic time-series observations started on December 02, 2023 at 01:03:18.92 UTC and ended December 02, 2023  at 12:09:33.18 UTC. During observations both NIRSpec detectors are read out, but light from WISE 0855 primarily falls on the first detector (NRS1). Based on JWST/NIRCam images of WISE 0855 from GTO program 1230 \citep{2017jwst.prop.1230A} no background sources were found at the location of WISE 0855 during our time-series observations.

\subsection{Stage 1 Detector Reduction}
We use the default reduction steps\footnote{\url{https://jwst-pipeline.readthedocs.io/en/latest/jwst/pipeline/calwebb_detector1.html}} for the Stage 1 detector processing, but describe each of these steps and their impact on the data in the following two sections. The first step of Stage 1 is \texttt{group\_scale}, which rescales groups in case in-flight averaging was done incorrectly before data downlink. Rescaling was not necessary for any of the segments. The \texttt{dq\_init} step initializes the data quality mask by pulling a reference pixel mask based on science observation metadata. The \texttt{saturation} step updates the mask by flagging pixels above the saturation threshold and below the response floor. The \texttt{superbias} step removes a reference bias image from each group in the ramp data. There is a 4-pixel wide, light insensitive region around the NIRSpec detectors used to estimate the contribution from readout electronics in the reference pixel step (\texttt{refpix}). 

There is odd-even row (alternating perpendicular to dispersion) pattern contribution that requires adding about 344 counts (odd row) or 350 counts (even row) to each group. Next, the \texttt{linearity} correction step is applied. The \texttt{dark\_current} step is skipped because there is no corresponding reference file appropriate for our science data to complete this step. The \texttt{jump} function removes outliers like cosmic rays detected while reading up the ramp. The Stage 1 detector pipeline reduction is completed by estimating slope images with the \texttt{ramp\_fitting} function and multiplying the slope images by the appropriate gain value determined by the detector readout in the \texttt{gain\_scale} step. The pipeline estimates the error associated with each pixel during the \texttt{ramp\_fitting} step and then updates the error array in the \texttt{gain\_scale} step.

\subsection{Stage 2 Spectroscopic Reduction}
 Stage 2 of the spectroscopic pipeline is used to create flux calibrated 2-d spectral images for aperture extraction. We use the default steps\footnote{\url{https://jwst-pipeline.readthedocs.io/en/latest/jwst/pipeline/calwebb_spec2.html}} for Stage 2 spectroscopic processing. The \texttt{assign\_wcs} step associates the world coordinate system with the science data and also imports the relevant wavelength map reference file. The \texttt{nsclean} function removes the residual 1/f noise caused by the readout electronics at the integration level for each slope image. The \texttt{extract\_2d} step clips the relevant spectral trace area from the full subarray image. The \texttt{flat\_field} corrects for the flat field and applies the absolute flux calibration to the 2-d spectral images. The \texttt{photom} step was run, but only pulls a reference file that multiplies the 2-d spectral images by 1.0\footnote{\url{https://jwst-docs.stsci.edu/jwst-calibration-status/nirspec-calibration-status/nirspec-calibration-concept}}. Pixels that are labeled as bad in the data quality mask are removed with the \texttt{pixel\_replace} function. The error image of each of the 2-d spectral images is updated once more during the flat fielding and flux calibration step. There are pixels within the error image that have \texttt{nan} values and these are replaced by doing a linear interpolation from surrounding pixels.

 \subsection{Spectral Extraction and Masking}
 Following the standard reduction steps described in the previous two subsections, we now implement a set of custom procedures required to properly center and extract the spectrum of WISE~0855. The center of the spectral trace is approximated by fitting a 1-dimensional Gaussian at every column along the spectral image for each pointing. The x and y positions are used to fit a second order Chebyshev series. Along this best fit trace center we extract a 4-pixel wide radius. The excess background is estimated by taking a median of the outer 3-pixels from both sides of the extraction aperture. The excess background is then subtracted from the extracted spectrum. Several different values of extraction radii and background were tested, but the chosen values maximized signal-to-noise and produced a spectrum that was most consistent with \cite{2024AJ....167....5L}. To estimate the error at every wavelength we apply the same 4-pixel extraction radius to the error image associated with each 2-d spectroscopic image.

\begin{figure}
    \centering
    \includegraphics[width=3.5in]{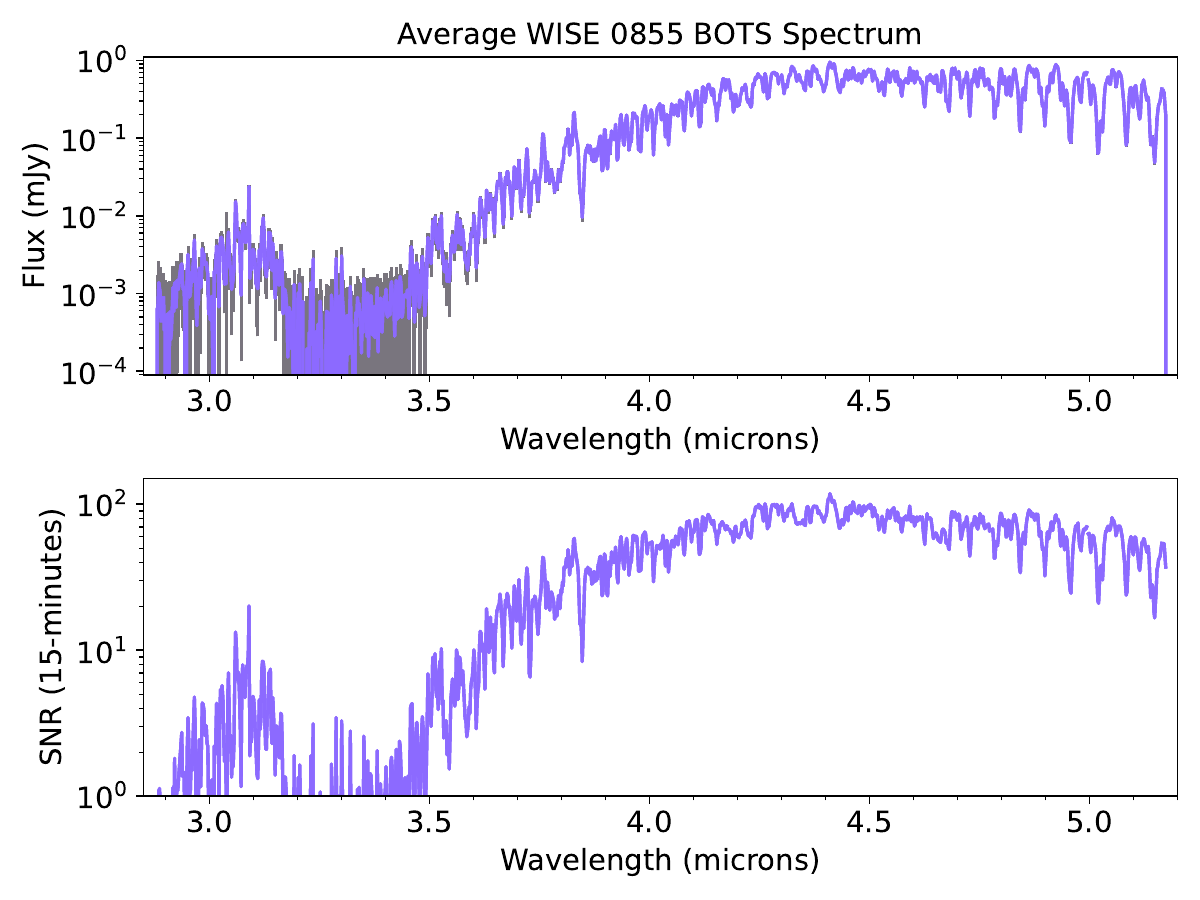}
    \caption{Top: The mean WISE 0855 spectrum over the 11-hour period (purple) in log flux scale. Representative error bars from the first 15-minute integration are plotted in gray. Bottom: signal-to-noise (SNR) of the spectrum per integration also in log scale. Signal-to-noise per integration remains above 5 at wavelengths longer than 3.6 $\mu$m. The average signal-to-noise between 3.827 - 5.170 $\mu$m is 73.}
    \label{fig:average_spectrum}
\end{figure}

Over the 11-hour time-series, WISE 0855 shifts downward in the spatial direction on the detector by $\sim$0.1 pixels based on the best fit traces. Variations in best fit trace between each 15-minute pointing changes the wavelength solution by, on average $1.8 * 10^{-6}$ $\mu$m, which is about 900 times smaller than the median difference between each spectral element $1.7 * 10^{-3}$ $\mu$m. After all 44 spectral extractions are done, the extracted spectra are over plotted and compared visually within 0.1 $\mu$m intervals. Individual pixels or pixel regions with systematically high or low flux values are masked out. This masking affects five regions. The pixel at 2.894 $\mu$m is masked because it has a flux value 2 times larger than surrounding wavelengths for all extracted spectra. The pixel at 4.178 $\mu$m is masked out because the spectrum varies by 200$\%$ and not in a physically smooth manner like nearby wavelengths. All fluxes at 4.692 $\mu$m are negative and therefore masked. All fluxes at 4.996 $\mu$m are relatively high compared to surrounding wavelengths. All fluxes at 5.166 $\mu$m are negative and masked. Single 1-2 pixel wide regions of hot pixels appear in seven spectra along the time-series: spectrum number 17, 28, 29, 30, 32, 37, and 44. An average spectrum of the reduced time-series data is plotted in Figure~\ref{fig:average_spectrum} with representative error bars of each 15-minute integration. These spectroscopic reductions differ from those published in \cite{2024ApJ...977L..49R} and \cite{2025arXiv251024575W}, respectively, in that we remove the excess background and use JWST-pipeline-derived errors for analysis.

\section{time-series Spectra and Light Curves}\label{sec:Features}
\begin{figure*}
    \centering
    \includegraphics[width=\textwidth]{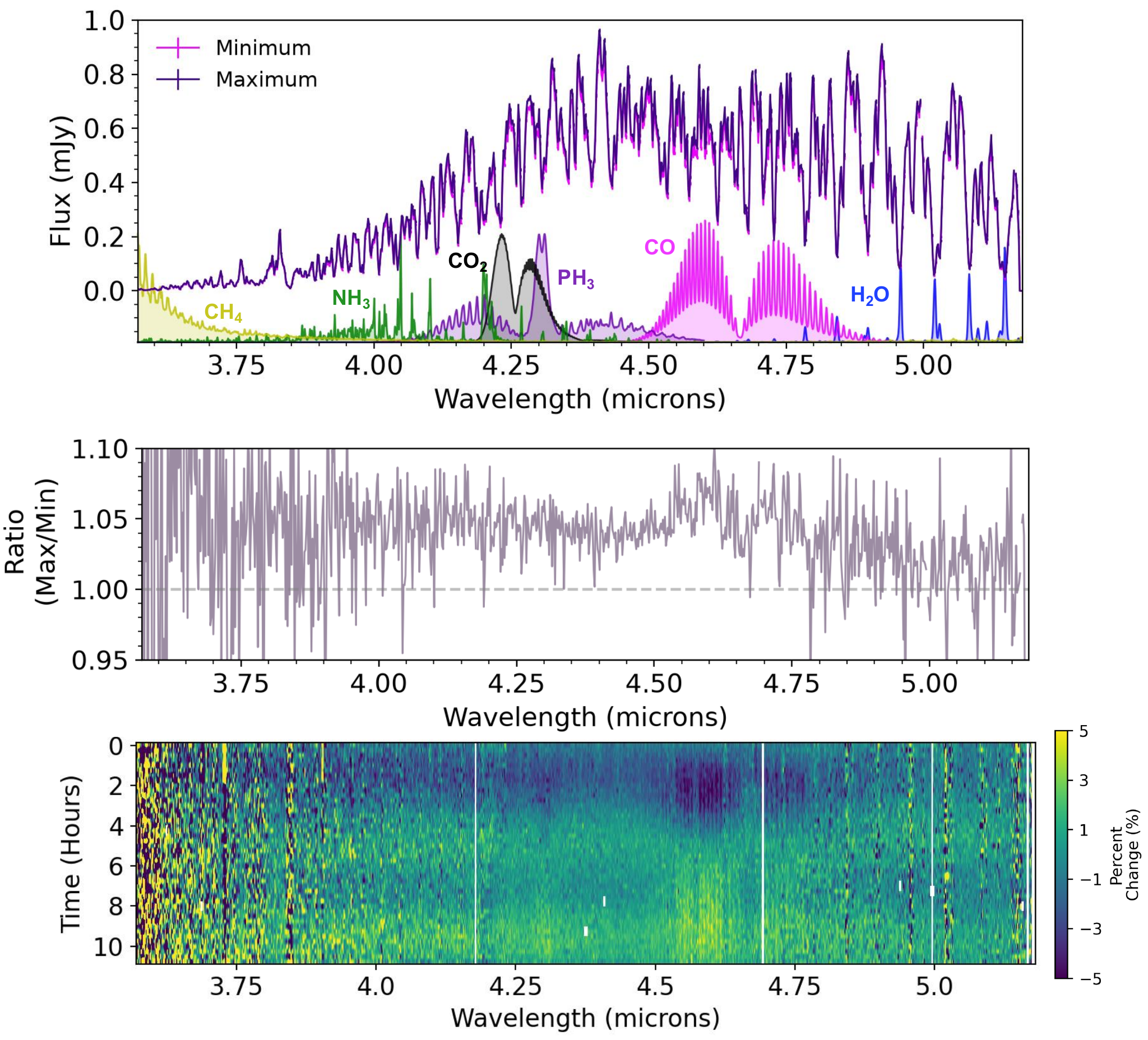}
    \caption{\textbf{Top:} Minimum (pink) and maximum (purple) brightness spectra with error bars plotted of the high signal-to-noise region of the time-series. \textbf{Middle:} Ratio of the maximum and minimum brightness spectra with a reference line plotted as a dashed gray line. \textbf{Bottom:} Image of the percent changes (blue-yellow) of the spectrum over time (hours). Both panels are truncated in wavelength compared to Figure~\ref{fig:average_spectrum}, due to the limited signal-to-noise at wavelengths shorter than 3.6 $\mu$m.}
    \label{fig:time_series}
\end{figure*}

\begin{figure*}
    \centering
    \includegraphics[width=7in]{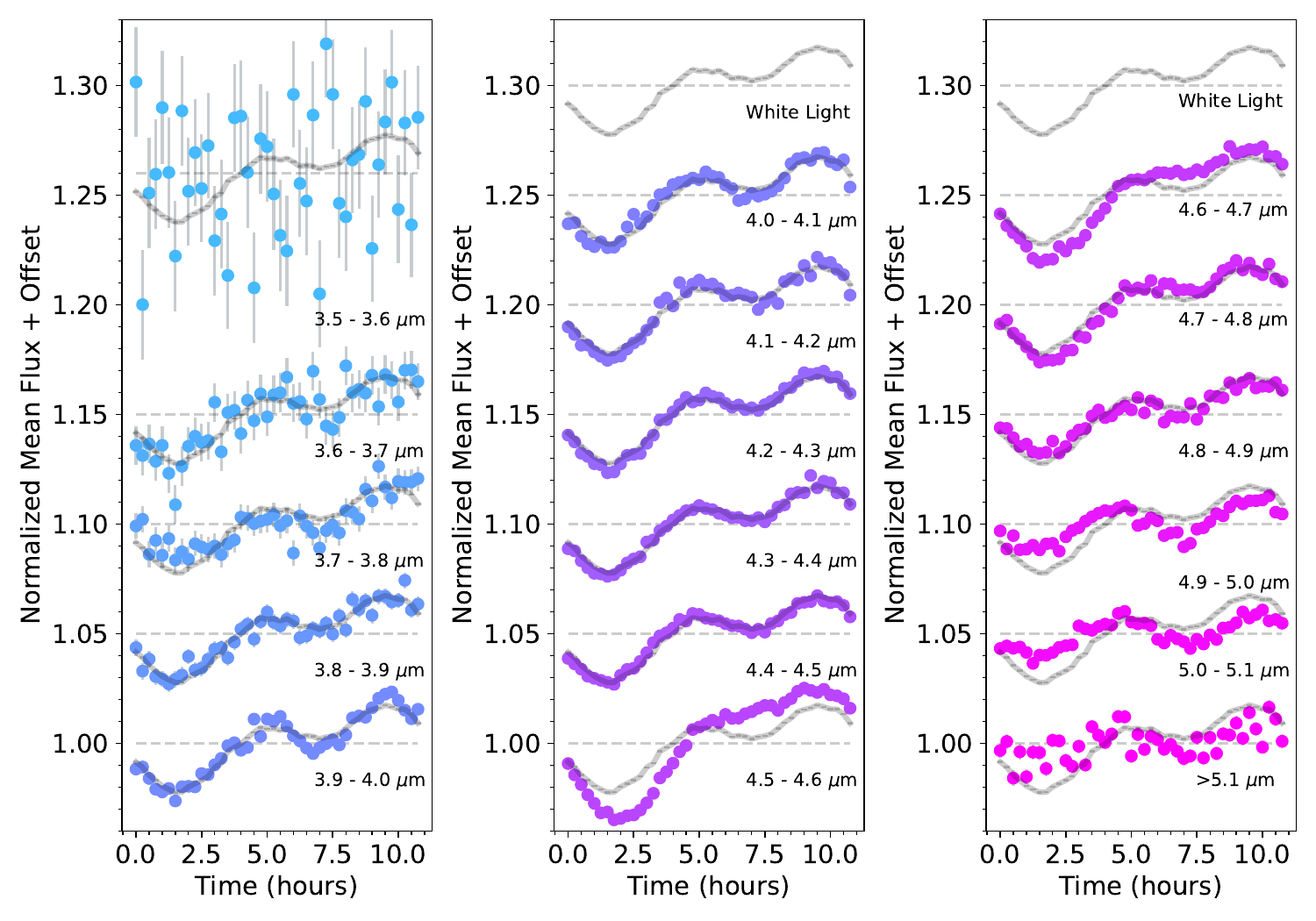}
    \caption{Normalized light curves of the NIRSpec data. Each colored light curve represents the mean over a wavelength bin width of 0.1~$\mu$m. Wavelengths shorter than 3.6 $\mu$m are low signal-to-noise. The white light curve represents the mean of the time-series over the entire wavelength range of the NIRSpec spectra. Each wavelength bin has the white light curve overplotted in gray to emphasize the relative differences as a function of wavelength. A reference line of no change is plotted as a gray dashed line over each light curve.}
    \label{fig:light curve}
\end{figure*}

The NIRSpec time-series spectra are visualized as both percent variation maps and light curves in this work. Each spectrum in the time-series is divided by the mean to calculate the percent variations over time. The percent variations along with the minimum and maximum brightness spectra are shown in Figure~\ref{fig:time_series}. Light curves are made by calculating the mean flux within bins of 0.1 $\mu$m width for each spectrum along the time-series (Figure~\ref{fig:light curve}). The largest amplitude spectral feature extends over 4.5 -- 4.8 $\mu$m. Lower amplitude, broad wavelength spectral changes are seen between 3.75 -- 5.1 $\mu$m. There is another variation near 4.25 $\mu$m that shows the same amplitude pattern as the 4.5 -- 4.8 $\mu$m feature. Light curves between 4.5 -- 4.8 $\mu$m have larger amplitudes, appearing dimmer at the $\sim$2 hour mark and brighter at the $\sim$7.5 hour mark compared to the white light curve. The longer wavelength light curves in the time-series beyond 4.8~$\mu$m show smaller amplitudes compared to the white light curve.  The light curves representing wavelengths shorter than 3.6~$\mu$m do not have enough signal-to-noise to capture significant variability. The light curves generally retain a similar shape across most wavelength bins. 

\subsection{Major Molecular Features}  
We visually search for variations that align with opacities of molecular species that have been detected or are predicted to appear in the atmospheres of Y-dwarfs. Retrieval analysis of WISE 0855 in \cite{2024ApJ...977L..49R} measured the atmospheric abundances of \ce{H2O}, \ce{CH4}, CO, \ce{CO2}, \ce{NH3}, and \ce{PH3}. Assuming a pressure of 1 bar and temperature of 260 K, we estimate the number density (n$_{\rm tot}$) of gas in a specific region of the atmosphere using the ideal gas law

\begin{equation}
    n_{\rm tot} = \frac{P}{k_{b} T}
\end{equation}
where P is the local pressure, T is the local temperature and k$_{b}$ is the Boltzmann constant. The total particle density is multiplied by the mole fraction ($\chi_{A}$) of a specific molecule to estimate the number density of that species (n$_{A}$), 

\begin{equation}
    n_{\rm A} = \chi_{\rm A} *  n_{\rm tot}.
\end{equation}
The molecular density is then multiplied by the opacity of that molecule to calculate the local opacity per volume for that molecule:

\begin{equation}
    \kappa_{\rm A} = \sigma_{\rm A} *  n_{\rm A}.
\end{equation}

The disequilibrium molecules have their opacities calculated at 260 K for simplicity, but their true opacities can be much higher depending on quench pressure. The opacities used are compiled in \cite{lupu_2022_6600976} which utilizes the following references for each molecule: CH$_{4}$: \citep{1992JChPh..97..773P,2013JMoSp.291...69Y,2014MNRAS.440.1649Y, 2021ApJ...920...85M}, CO: \citep{2010JQSRT.111.2139R, GORDON20173, Li_2015, 2021ApJ...920...85M}, CO$_{2}$: \citep{2014JQSRT.147..134H, 2021ApJ...920...85M}, H$_{2}$O: \citep{2018MNRAS.480.2597P, 2017JQSRT.187..453B}, NH$_{3}$: \citep{Yurchenko_2011, 2016JQSRT.168..193W, 2021ApJ...920...85M}, PH$_{3}$: \citep{10.1093/mnras/stu2246, 2021ApJ...920...85M}.

\begin{figure*}
    \centering
    \includegraphics[width=\textwidth]{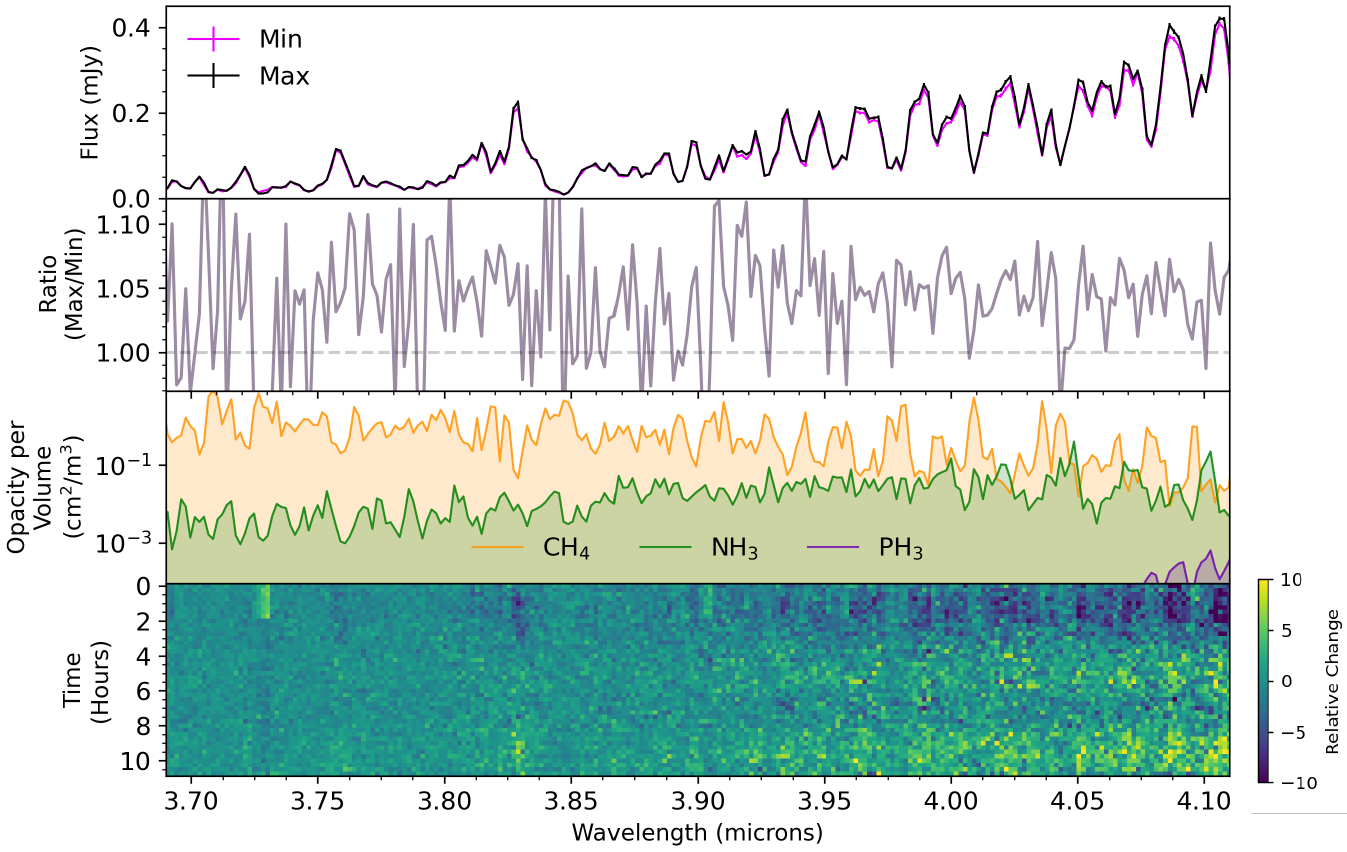}
    \caption{\textbf{Top}: Minimum (pink) and maximum (black) spectra plotted with error bars from 3.688 -- 4.263 $\mu$m. The error bars are so small that they are not visible. \textbf{Upper Middle}: Ratio of the maximum and minimum brightness spectra. \textbf{Lower Middle}: Opacities of methane (yellow), ammonia (green), carbon dioxide (black), and phosphine (purple) at 1 bar and T~=~260~K. \textbf{Bottom:} Relative change (blue-yellow) of the spectra over time (hours). This plot is similar to the heat map in Figure~\ref{fig:time_series}, but each wavelength element is divided by the white light curve shown in Figure~\ref{fig:light curve}. Regions where the opacity of methane and ammonia are smaller, stronger variations appear.}
    \label{fig:features1}
\end{figure*}

\begin{figure*}
    \centering
    \includegraphics[width=\textwidth]{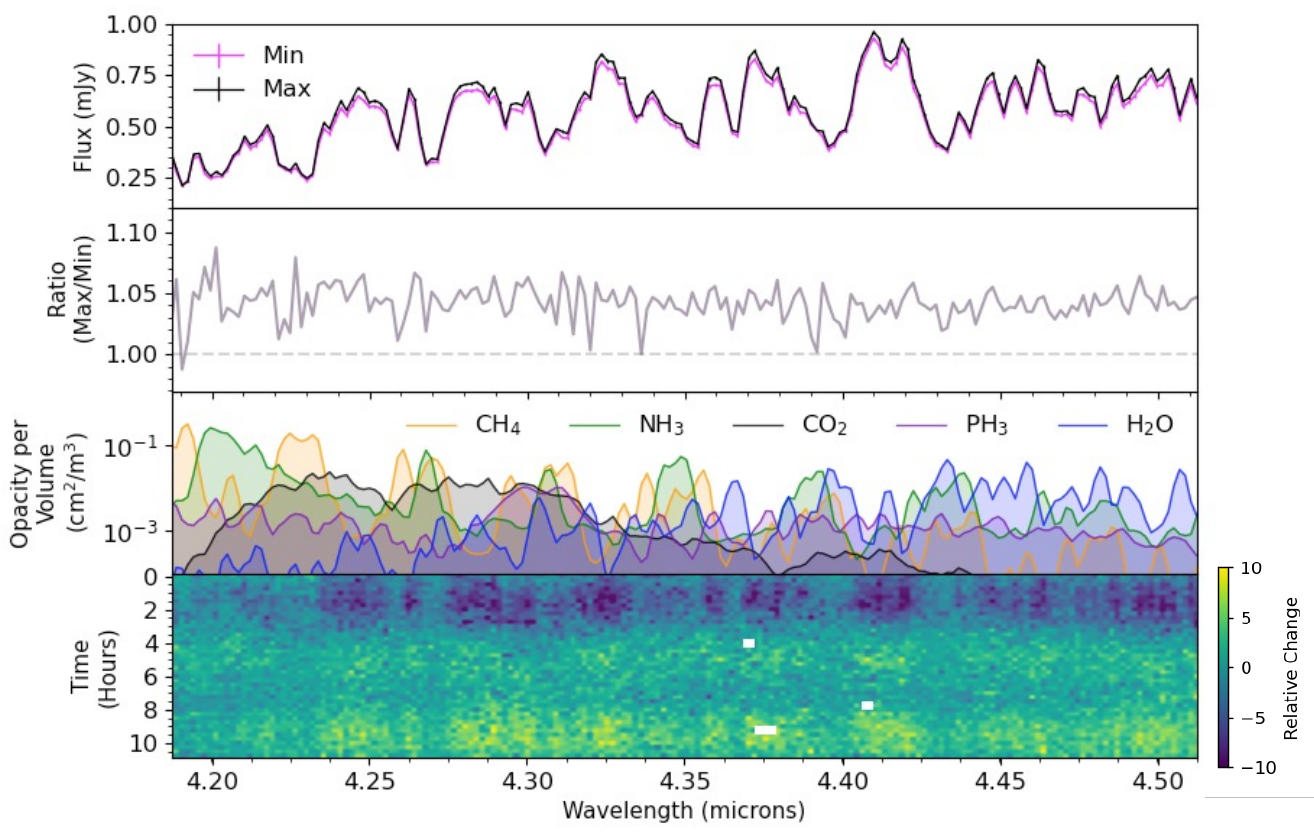}
    \caption{Same set up as Figure~\ref{fig:features1} but wavelength coverage is 4.188 -- 4.513 $\mu$m. The opacities of \ce{CO2} (black) and \ce{H2O} (blue) are added. Where the opacity of ammonia and methane are low, wavelengths that have strong contributions from \ce{CO2} and \ce{PH3} have stronger relative variations. This suggests the variations are driven by apparent abundance changes for these molecules.}
    \label{fig:features2}
\end{figure*}

\begin{figure*}
    \centering
    \includegraphics[width=\textwidth]{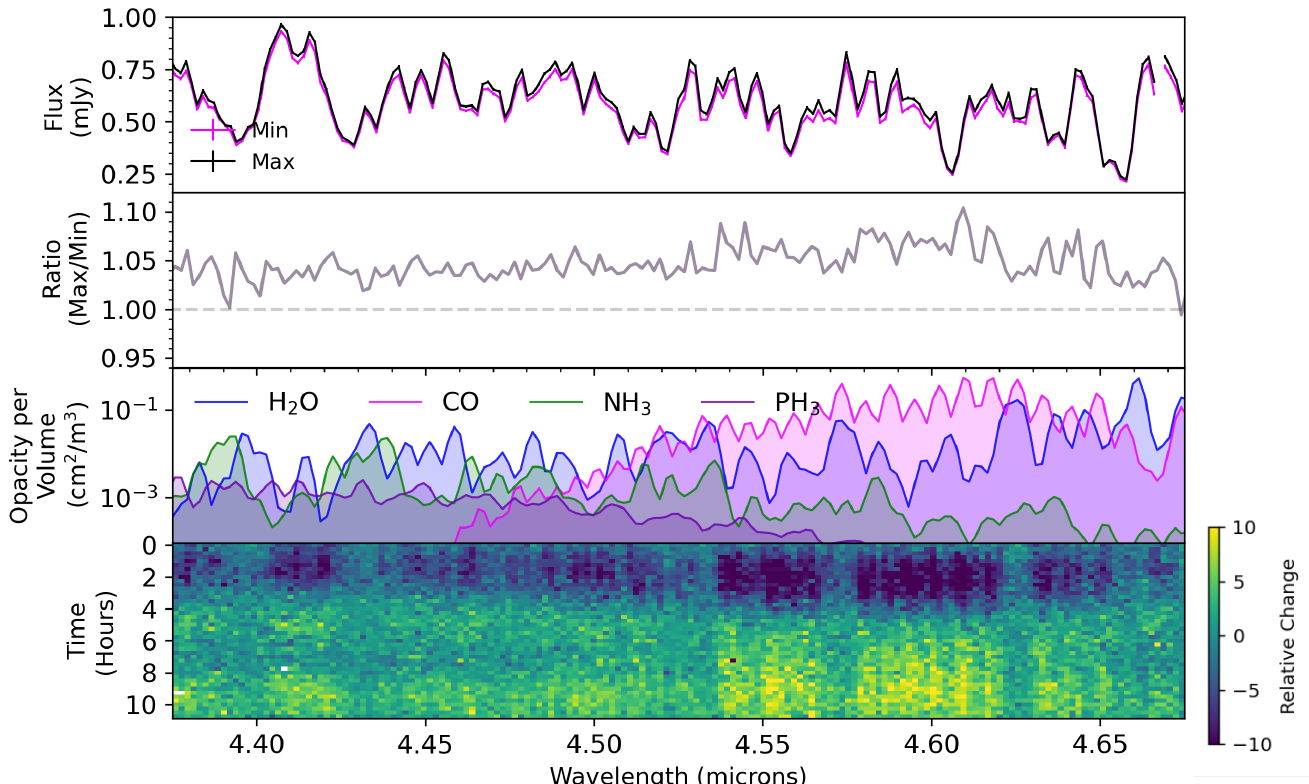}
    \caption{Same set up as Figure~\ref{fig:features1} but wavelength coverage is from 4.375 -- 4.675 $\mu$m. Wavelengths with significant opacity from carbon monoxide contain the highest amplitude changes within the time-series. This implies changes in CO abundance contribute the most to WISE 0855's observed variability.}
    \label{fig:features3}
\end{figure*}

\begin{figure*}
    \centering
    \includegraphics[width=\textwidth]{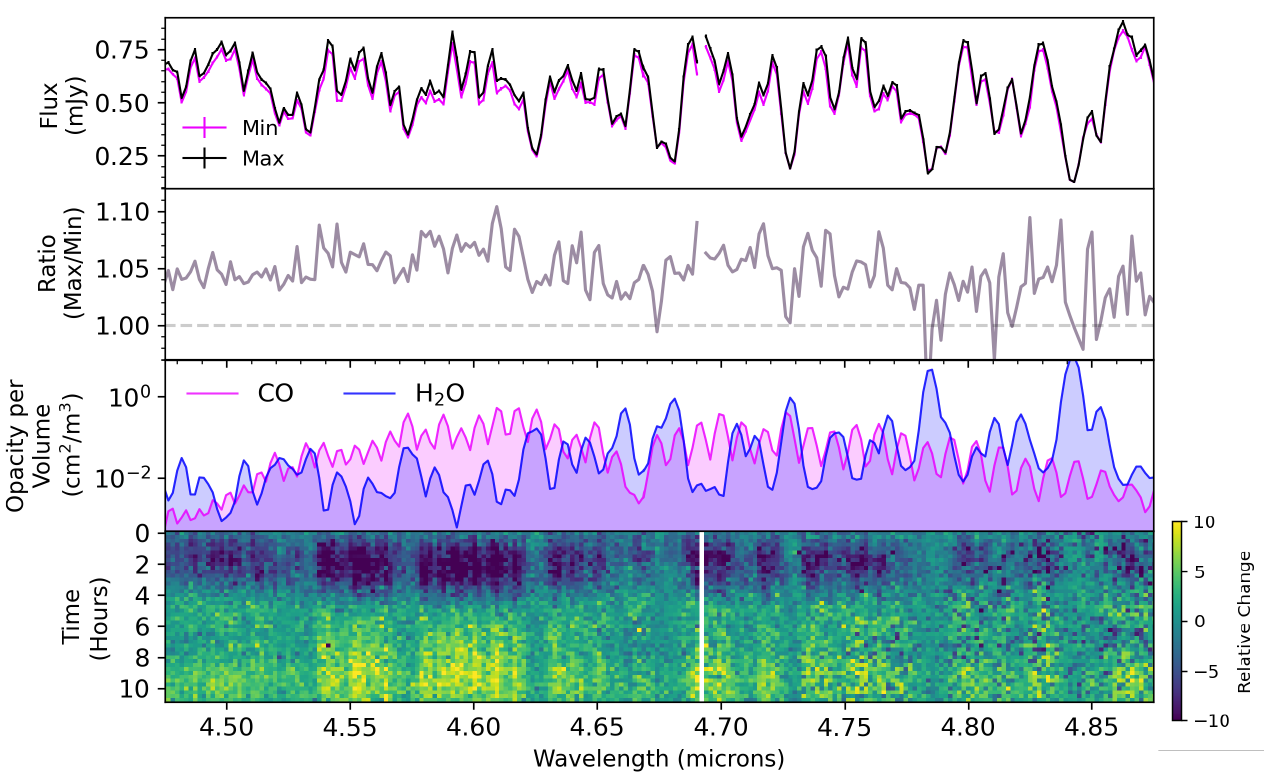}
    \caption{Same set up as Figure~\ref{fig:features1} but wavelength coverage is from 4.475 -- 4.875 $\mu$m. Wavelengths where water vapor has a higher opacity than carbon monoxide show less relative variations.}
    \label{fig:features4}
\end{figure*}

\begin{figure*}
    \centering
    \includegraphics[width=\textwidth]{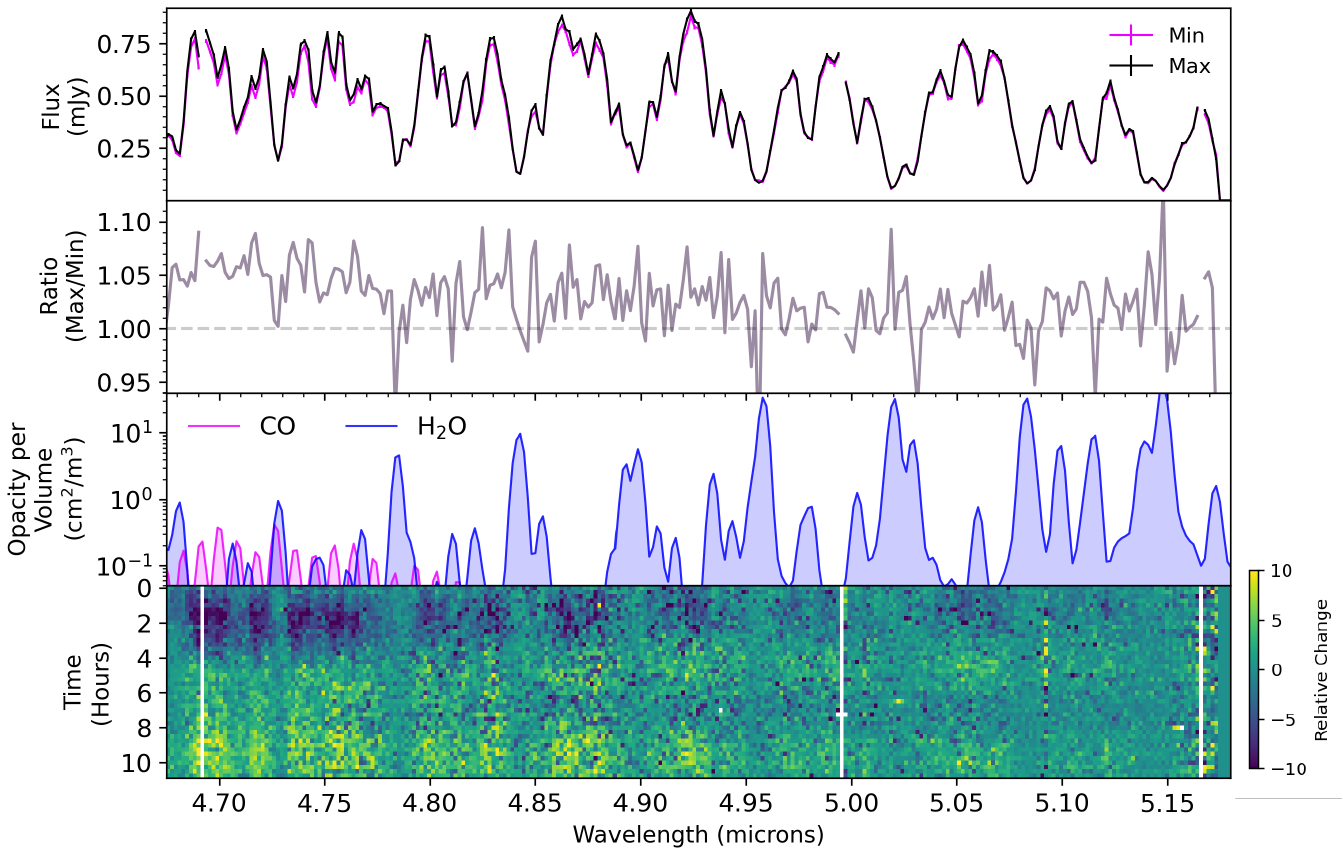}
    \caption{Same set up as Figure~\ref{fig:features1} but wavelength coverage is from 4.675 -- 5.18 $\mu$m. There are relative variations between water vapor lines that do not line up with any expected molecule. This suggests that some of WISE 0855's variability may not be the result of changes in molecular abundances.}
    \label{fig:features5}
\end{figure*}

In Figures~\ref{fig:features1} -- \ref{fig:features5}, we plot estimated molecular opacities alongside the minimum and maximum brightness spectra with maps of the percent variability relative to the white light curve of WISE 0855. \ce{CH4} and \ce{NH3} have significant opacities between 3.7 - 4.25 $\mu$m and 4.05 - 4.5 $\mu$m, respectively. Wavelengths with significant contributions from \ce{CH4} and \ce{NH3} show less variability compared to the averaged light curve (Figure~\ref{fig:features1}). In regions where the opacity of both \ce{CH4} and \ce{NH3} are minimal, relative variations are larger and the difference between the minimum and maximum brightness spectrum are greatest. Within the 4.2 - 4.5 $\mu$m region there are strong variations that coincide with opacity peaks of \ce{CO2} and \ce{PH3} in between the low opacity regions of \ce{CH4}, \ce{NH3}, and \ce{H2O} (Figure~\ref{fig:features2} and \ref{fig:features3}). The largest amplitude variations in the time-series coincide with the opacity of \ce{CO} from 4.5 - 4.8 $\mu$m (Figure~\ref{fig:features4}). 

The opacity of \ce{H2O} has several peaks from 4.5 - 5.18~$\mu$m which also overlaps the \ce{CO} absorption bands. The relative variations are lower where \ce{H2O} has bigger opacity values. There are some wavelengths where the opacity per volume of \ce{H2O} is lower than \ce{CO} which have a low relative variation. This is partially due to the assumption of one temperature and pressure for the entire spectrum and the sensitivity of these molecules to the assumed pressure temperature profile. In regions where \ce{H2O} has a low opacity but no other significant molecular opacities ($>$ 4.8 $\mu$m), there are stronger variations similar to the changes seen in between the \ce{CH4} and \ce{NH3} bands (Figure~\ref{fig:features5}). Many of the modulations in the time-series align with opacities of expected molecular gases; however, some changes are likely not explicitly related changes in molecular abundances. We explore the origin of changes within the time-series spectra in more detail by comparing the data to atmospheric models in later sections.

\section{Analysis} \label{sec:Analysis}

\subsection{Principal Component Analysis}
\begin{figure*}
    \centering
    \includegraphics[width=7in]{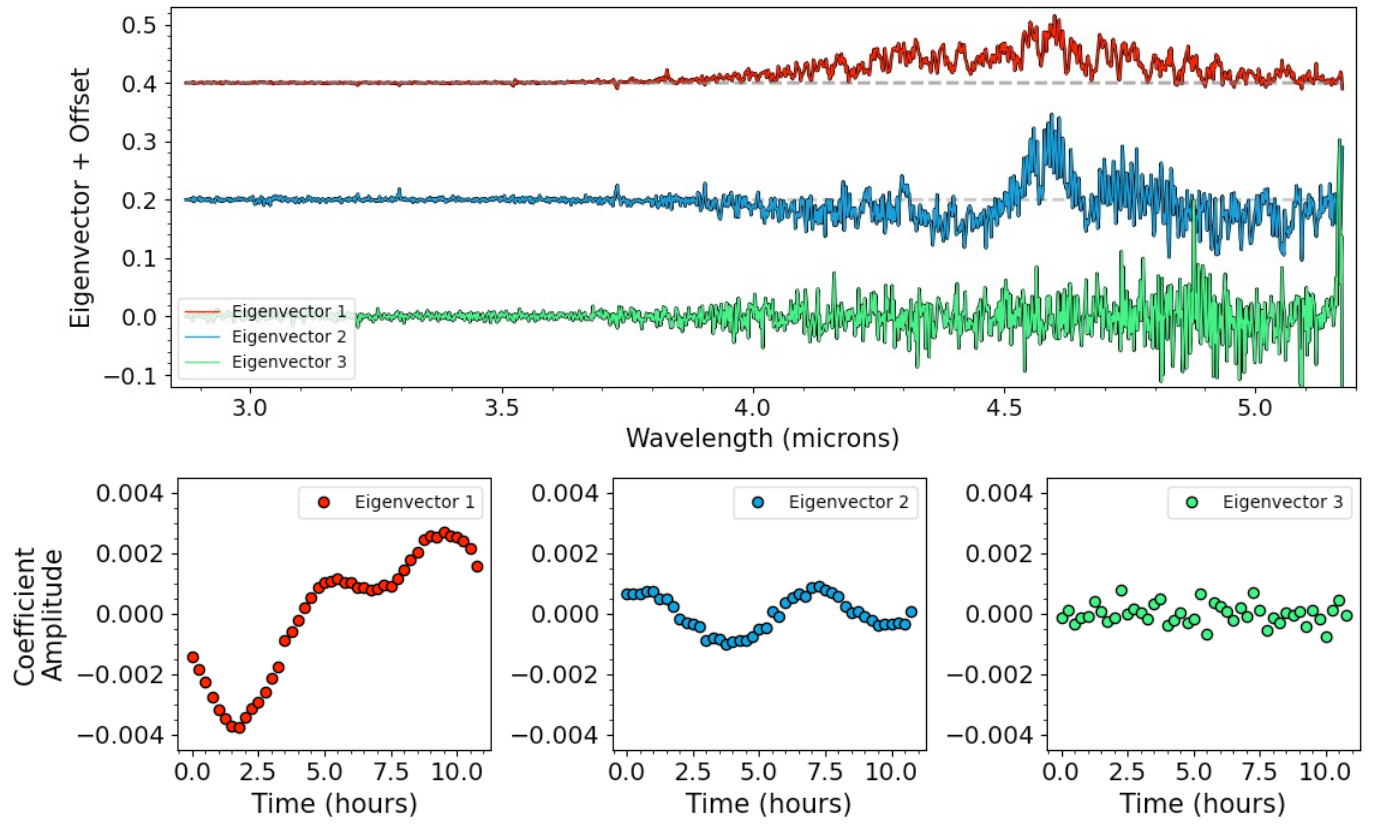}
    \caption{\textbf{Top:} First three eigenspectra of the WISE 0855 spectral time-series. Each eigenspectrum is offset by 0.2 for visual clarity. A dashed light-gray line is plotted with each eigenspectrum to show the zeropoint. \textbf{Bottom:} Best fit coefficient for each eigenvector over time. The first eigenvector (red) has the largest amplitude change. The amplitude and period of the second eigenvector (blue) has a different shape and period than the first.}
    \label{fig:PCA}
\end{figure*}

\begin{figure}
    \centering
    \includegraphics[width=3.5in]{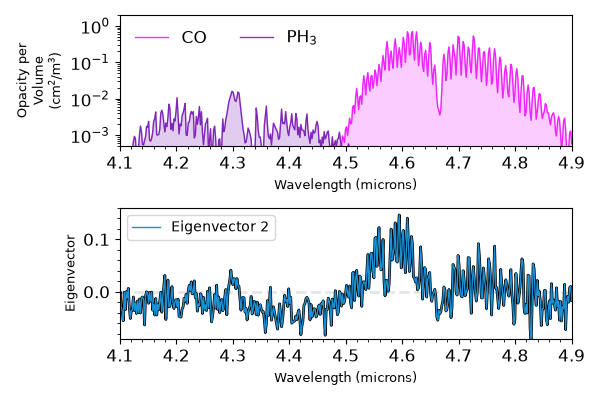}
    \caption{Zoomed panel of Eigenvector 2 from Figure~\ref{fig:PCA}. \textbf{Top:} Opacities of disequilibrium molecules at T$_{\rm eff}$ = 200 K and 1 bar. \textbf{Bottom:} The second eigenvector has strong contributions where CO and PH$_{3}$ are prominent. }
    \label{fig:PCA opacity}
\end{figure}

We use principal component analysis (PCA) to understand the dominant sources of variability within the time-series spectra. Each spectrum is divided by the sum of the entire time-series to normalize the entire set to one. The \texttt{PCA} function from the \texttt{scikit-learn} package \citep{scikit-learn} is used to find a set of eigenspectra and eigenvalues that form the basis of our data set. The signature of a single eigenspectrum can represent the modulation of multiple atmospheric parameters and each eigenspectrum is interpreted as distinct contributions from unresolved surface features \citep{2026arXiv260726182S}. As WISE 0855 rotates, the amplitude or coefficients of each eigenspectrum changes over time. The first and second largest eigenvalues contribute to 60.5\% and 4.7\% of the variability, respectively. All other eigenvectors contribute 1.5\% or less variability individually, however 34.7\% of the total variability is not captured by the dominant two eigenvalues implying complex behavior in the time-series. We use a minimizing function to determine the coefficients for each eigenvector required to recreate a single spectrum at every point along the time-series. The first three eigenspectra and their best fit coefficients over time are shown in Figure~\ref{fig:PCA}. The first and most dominant eigenvector displays broad wavelength changes that are similar in shape to the mean spectrum. The second eigenvector has the strongest response within wavelengths impacted by disequilibrium molecules like \ce{CO} and \ce{PH3} (Figure~\ref{fig:PCA opacity}). The third eigenvector is relatively noisy and the best fit coefficients are scattered around zero unlike the first two eigenvectors. This implies the variations from the first two eigenvectors are primarily astrophysical and that all other eigenvectors are capturing the scatter from measurement noise. The coefficients of eigenvector 1 and 2 have different shapes over time suggesting the variations originate from different surface features on WISE 0855.

\subsection{Periodicity of WISE 0855 and Long-Term Variability}
\begin{figure*}
    \centering
    \includegraphics[width=7in]{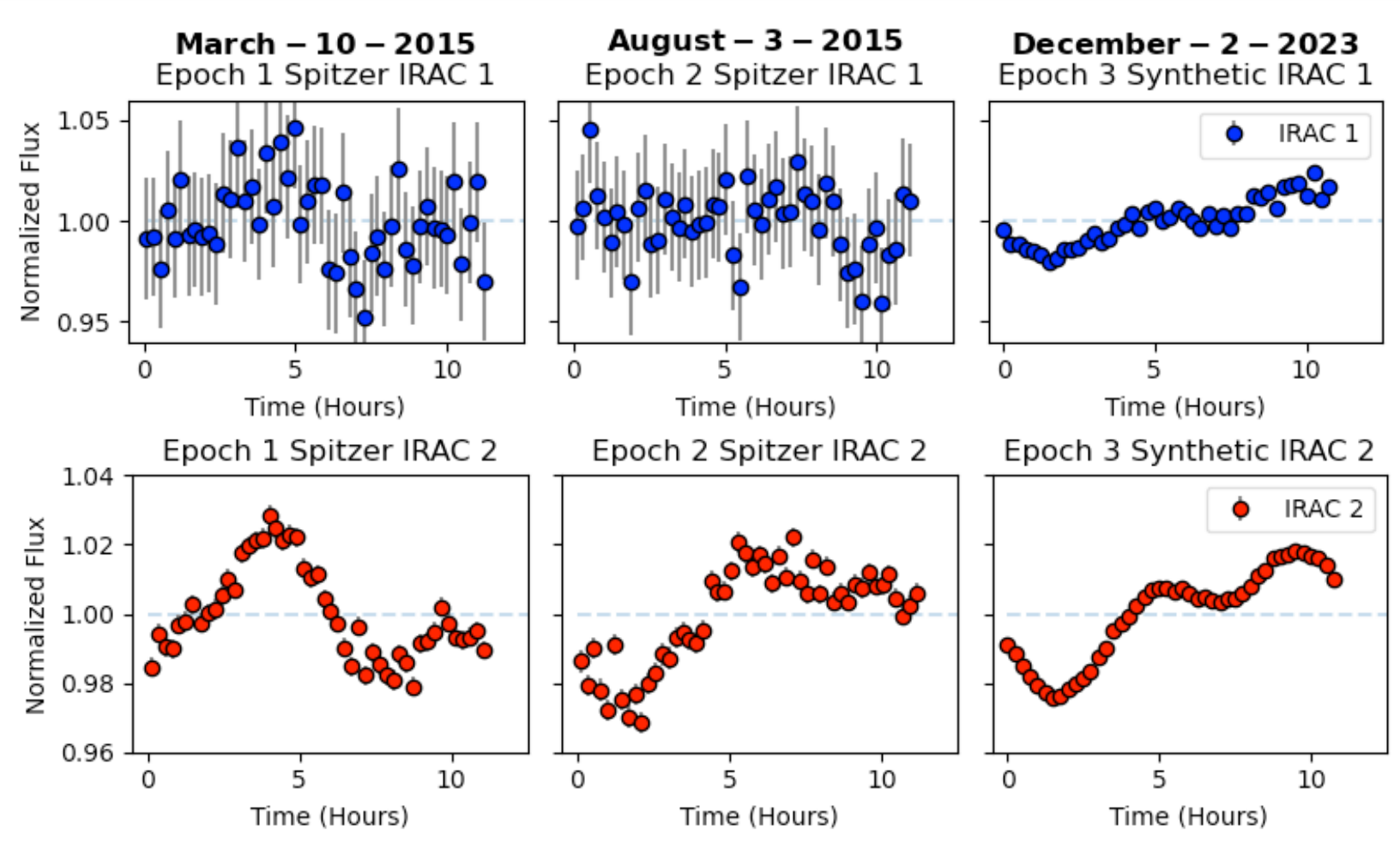}
    \caption{time-series photometry of WISE 0855 over three epochs within the Spizter IRAC 1 (blue) and IRAC 2 (red) bands. The data from \cite{2016ApJ...832...58E} is binned to 15-minute intervals to match the JWST cadence. Epoch 3 is the JWST/NIRSpec data convolved with the Spitzer photometric bands.}
    \label{fig:multi epoch photometry}
\end{figure*}

\begin{figure*}
    \centering
    \includegraphics[width=7in]{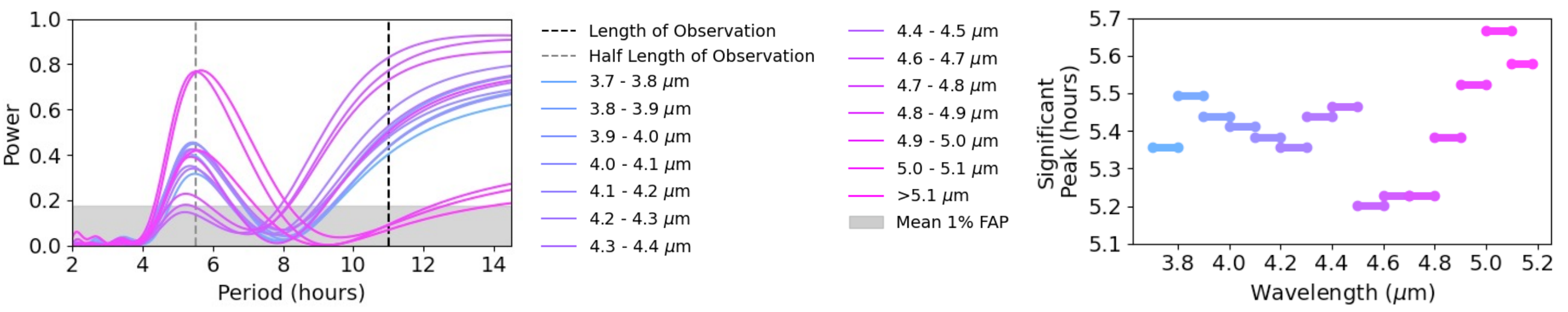}
    \caption{\textbf{Left:} Periodogram of the light curves from Figure~\ref{fig:light curve}. The color of each period distribution corresponds to the relevant wavelength bin. The vertical dashed lines represent the total time of the observation (black) and half of that time (gray). The mean 1\% false alarm probability is plotted horizontally in gray. \textbf{Right:} Only considering peaks shorter than the 11 hour observation duration, the time (in hours) at which the peak amplitude occurs is plotted versus wavelength bin. }
    \label{fig:LS1}
\end{figure*}

Previous studies of WISE 0855 have not determined the rotation period of the object, and no inclination has been estimated to date. In \cite{2025arXiv251024575W}, portions of WISE 0855's light curve were fit using a gaussian process model using a quasi-periodic kernel suggesting periods longer than 14 hours for the object. We also look for significant periods in the binned light curves from Figure~\ref{fig:light curve} using the \texttt{LombScargle} function from astropy \citep{astropy:2013, astropy:2018, astropy:2022}. The 1\% false alarm probability is estimated using the same function with the bootstrap method. For wavelength bins longer than 3.7 $\mu$m there is significant power for periods between 5.4 - 5.6 hours however, longer periods are also likely (Figure~\ref{fig:LS1}) and no significant periods are found in the time-series when 75$\%$ or less of the 11 hour observation is used. We place our JWST/NIRSpec data in context to the time-series photometry published in \cite{2016ApJ...832...58E} by convolving the data set with the Spitzer IRAC1 and IRAC2 filters (Figure~\ref{fig:multi epoch photometry}). In the work by \cite{2016ApJ...832...58E}, WISE 0855 was observed on two different epochs, 2015 March 10 and 2015 August 3. WISE 0855 was observed first by IRAC 2 for 11.3 hours at a cadence of 96.8 seconds then with IRAC 1 at a cadence of 93.6 seconds for both epochs. A 3$\sigma$ clip is applied to the photometric time-series to remove outliers and the data are binned to a cadence of 15-minutes. 

\begin{figure*}
    \centering
    \includegraphics[width=7in]{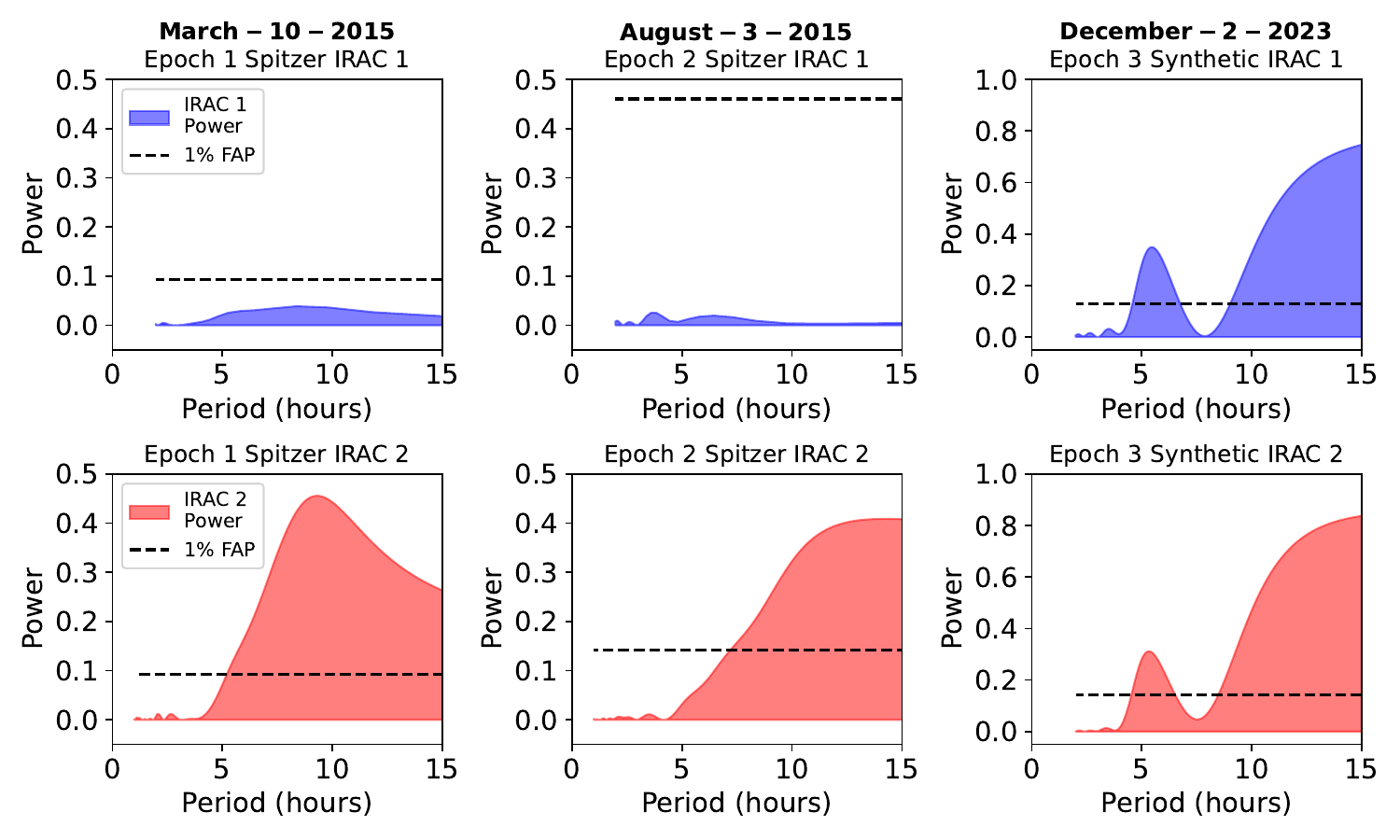}
    \caption{Periodograms of WISE 0855 time-series photometry at three different epochs. The power distribution of the IRAC~1 observations are shown at the top and IRAC2 observations on the bottom row. The power distribution of the periods are shown in color with the 1\% false alarm probability of the peak shown as a black dashed line. Epoch 1 and  2 data are from \cite{2016ApJ...832...58E}. The Epoch 1 observations in the IRAC 2 bandpass have a significant peak at 9.33 hours. Significant peaks at 5.4 hours (IRAC 1) and 5.5 hours (IRAC 2) are seen during Epoch 3. In all three cases there is meaningful contribution of periods longer than the time used on any single observation.}
    \label{fig:LS2}
\end{figure*}

The synthetic IRAC~1 photometry derived from JWST/NIRSpec is on average 37$\%$ fainter than the measured IRAC~1 photometry. This offset between the synthetic and measured IRAC~1 photometry has been observed in other cold brown dwarfs \citep{2024AJ....167....5L, 2024ApJ...973..107B,2025ApJ...993..165S}. Work by \cite{2024ApJ...973..107B} compared the photometric offsets between brown dwarfs and calibration stars suggesting that the photometric offset is caused by a light leak in JWST/NIRSpec primarily impacting redder objects. The synthetic IRAC 2 photometry is on average 7$\%$ fainter than the measured IRAC~2 data, but within the observed scatter in \cite{2024ApJ...973..107B}. The synthetic photometry of the JWST data has similar peak-to-peak amplitudes (4 -- 6\%) to the previously published Spitzer photometry. Both epochs of the IRAC 1 photometry have no significant periods above the false alarm probability. Epoch 1 of the IRAC~2 photometry has a significant period at 9.33 hours, whereas epoch 2 only has significance for periods longer than the duration of the observations.  Both epochs of the IRAC 2 photometry have significant power for periods longer than the length of observations. The periodograms of all epochs are shown in Figure~\ref{fig:LS2}.

WISE 0855 has shown amplitude modulations on timescales between $\sim$5 and 10 hours, but the exact rotation period of WISE 0855 is still unknown. The periods with significant power in the NIRSpec/JWST light curves are about half of the time-series duration. Longer baseline observations ($>$20 hours) will be required to determine if the periodicity shows evolution or contains multiple components due to rotation, zonal winds, or large scale planetary waves as seen in other brown dwarfs \citep{2017Sci...357..683A,2021ApJ...906...64A}.

\section{Atmospheric Model Fitting} \label{sec:Model Fitting}
\subsection{Models}
In this section, we compare different atmospheric model grids to the time-averaged spectrum and each spectrum along the JWST/NIRSpec time-series. Several temporal changes were linked to molecular gases such as \ce{CO, PH3, and CO2} in Section~\ref{sec:Features}. WISE 0855's upper atmosphere is cold enough to condense \ce{H2O} clouds based on theoretical models and previous work has shown WISE 0855's molecular features are impacted by vertical convective transport. We try to quantify the impact of convection-driven disequilibrium chemistry and \ce{H2O} clouds using published and custom self-consistent atmospheric models including these effects. For all models we assume surface gravity and metallicity are fixed at log(g)~=~4 and [M/H]~=~0, because these are bulk properties which are unlikely to change over our observations.

The first set of self-consistent atmospheric model grids used were created using PICASO \citep{2023ApJ...942...71M} and \texttt{Virga} \citep{2020zndo...3759888B} to include the contribution of water clouds \citep{2026ApJ..1000...98M}. \texttt{Virga} was used to include the contribution of water clouds within PICASO based on the \cite{2001ApJ...556..872A} f$_{\rm sed}$ cloud prescription. The second set of atmospheric model grids are updated coolTLUSTY models first presented in (\citealt{2023ApJ...950....8L}, Lacy et al. submitted) with new \ce{CH4} opacities from \cite{2020ApJS..247...55H}, higher spectral resolution, and \ce{CO2} included. The PICASO and coolTLUSTY disequilibrium atmospheric models have updated quench chemistry treatment for CO$_{2}$ \citep{2014ApJ...797...41Z,2025RNAAS...9..108W, 2026ApJ..1000...98M}. A summary of all atmospheric model grids and input parameters are listed in Table~\ref{tbl:models}. We describe each of the grids in the following paragraphs.

The first grid of PICASO models are cloudless and do not include the impact of disequilibrium chemistry. This grid spans 250 -- 275 K in effective temperature, with 5 K spacing. Three derivative grids for the cloudless models were created where the abundances of CO, \ce{CO2}, and \ce{PH3} are individually post processed to have uniform volume mixing ratios throughout the entire P-T profile. The CO mini grid is calculated at log volume mixing ratio (VMR) values of -5 to -7.5 in 0.5 dex increments. The \ce{CO2} mini grid is calculated at log VMR values of -8.5 to -10 in 0.5 dex increments. The \ce{PH3} mini grid is also calculated at log VMR values of -8.5 to -10 in 0.5 dex increments. The fourth mini grid fixes the log VMR of CO at -7.0 while changing the \ce{CO2} and \ce{PH3} abundances simultaneously. The log VMR of \ce{CO2} and \ce{PH3} are both individually adjusted from -7.5 to -9.0 in 0.5 dex increments. A second cloudless PICASO grid includes disequilibrium chemistry parameterized by the eddy diffusion coefficient (K$_{\rm zz}$) and the mixing lengthscale is assumed to be the pressure scale height (H). A K$_{\rm zz}$ of 10$^{7}$ $\rm \frac{cm^{2}}{s}$ is assumed for all models based on the estimates derived from \ce{CO} in \cite{2020AJ....160...63M}. Four additional derivative disequilibrium grids are created by post-processing the abundances of CO, CO$_{2}$, and PH$_{3}$ the same way it was done for the equilibrium PICASO.

The next two major PICASO grids include water clouds using \texttt{VIRGA} with the $\rm f_{\rm sed}$, sedimentation efficiency parameter, from \cite{2001ApJ...556..872A} where a larger $\rm f_{\rm sed}$ corresponds to vertically compressed and optically thicker clouds, whereas smaller  $\rm f_{\rm sed}$ values produce more vertically extended and optically thicker clouds. With the exception of the mini grids where CO is fixed and CO$_{2}$ and PH$_{3}$ change, all cloudy models are calculated at f$_{\rm sed}$ values of 4, 6, and 8. The cloudy equilibrium PICASO model grid covers effective temperatures from 250 -- 275 K, with 5 K spacing. The disequilibrium, cloudy models have the same effective temperature range and spacing, but are calculated with a K$_{\rm zz}$ of 10$^{7}$ $\rm \frac{cm^{2}}{s}$. Identical to the cloudless models, derivative grids adjusting CO, CO$_{2}$, and PH$_{3}$ were run for the equilibrium and cloudy disequilibrium models.

The cloudfree, equilibrium coolTLUSTY model grid spans 200 -- 450 K in effective temperature, with 25 K spacing. Another version of the cloudfree coolTLUSTY grid includes disequilibrium chemistry assuming K$_{\rm zz}$ of 10$^{6}$ $\rm \frac{cm^{2}}{s}$ in radiative layers and mixing of $\sim$10$^{7}$-10$^{9}$ $\rm \frac{cm^{2}}{s}$ in convective layers following mixing length theory with a mixing length of one pressure scale height. It covers the same effective temperature range as the cloudfree, equilibrium grid. Another cloudfree, disequilibrium grid maintains the K$_{\rm zz}$ = 10$^{6}$ $\rm \frac{cm^{2}}{s}$ assumption in radiative layers, but then adopts different mixing length scale values in convective layers: L = H, L = 0.1 x H, L = 0.01 H, where H is the pressure scale height. At WISE-0855's surface gravity and temperature, chemical quench points all land in convective layers for these models.

The cloudy coolTLUSTY models parameterize water clouds using the variable CDAMPU\footnote{Note that in \cite{2023ApJ...950....8L} equation (2) this parameter was referred to as TCUP, inconsistent with prior coolTLUSTY-based publications. Here we return to the more standard nomenclature of CDAMPU.} from \cite{2023ApJ...950....8L}, calculated at values of 2 and 6. CDAMPU sets the ratio of the gaseous materials' pressure scale height to the cloud particles' pressure scale height. CDAMPU also controls how quickly the water cloud tapers off, with higher values of CDAMPU corresponding to less vertically extended clouds. The effective temperature range of the cloudy models is 200 -- 350 K with 25 K spacing. An equilibrium cloudy grid and disequilibrium cloudy grids were calculated.

\begin{table*}
\tiny
\centering
\hspace*{-1.2in}\begin{tabular}{l|cllcc}
 \textbf{Grid Description }           & T$_{\rm eff}$ Range & Cloud               &  DisEQ                &   Mixing         & Adjusted Species            \\ 
                                      &    (K)              & Parameter$^{a}$     &  Parameter$^{b}$       &   Lengthscale$^{c}$  &log(VMR)$^{d}$             \\\hline
\multicolumn{6}{l}{\textbf{PICASO - Clear, Equilibrium}} \\ \hline                                                               
Base Grid                                      & 250 -- 275    & --           &  --                   &       --          & -                    \\  
 Adjusted CO                                    & 250 -- 275    & --           &  --                   &       --          & CO: -7.5 -- -5       \\ 
 Adjusted CO$_{2}$                              & 250 -- 275    & --           &  --                   &       --          & CO$_{2}$: -10 -- -8.5\\ 
 Adjusted PH$_{3}$                              & 250 -- 275    & --           &  --                   &       --          & PH$_{3}$: -10 -- -8.5\\  
 Adjusted CO$_{2}$, PH$_{3}$, fixed CO           & 250 -- 275    & --           &  --                   &       --          & CO = -7, CO$_{2}$: -9 -- -7.5, PH$_{3}$: -9 -- -7.5  \\  \hline
\multicolumn{6}{l}{\textbf{PICASO - Clear, Disequilibrium}}  \\ \hline   
 Base Grid*                                      & 150 -- 275    & --           &  K$_{\rm zz}$ = 7         &       H          &          -            \\ 
  Adjusted CO                                    & 250 -- 275    & --           &  K$_{\rm zz}$ = 7         &       H          & CO: -7.5 -- -5       \\ 
  Adjusted CO$_{2}$                              & 250 -- 275    & --           &  K$_{\rm zz}$ = 7        &       H          & CO$_{2}$: -10 -- -8.5\\ 
  Adjusted PH$_{3}$                              & 250 -- 275    & --           &  K$_{\rm zz}$ = 7         &       H          & PH$_{3}$: -10 -- -8.5\\ 
  Adjusted CO$_{2}$, PH$_{3}$, fixed CO           & 250 -- 275    & --           &  K$_{\rm zz}$ = 7          &       H          & CO = -7, CO$_{2}$: -9 -- -7.5, PH$_{3}$: -9 -- -7.5  \\ \hline

 \multicolumn{6}{l}{\textbf{PICASO - Cloudy, Equilibrium}}  \\ \hline   
  Base Grid                                     & 250 -- 275    & f$_{\rm sed}$ = 4, 6, 8 &  --                    &       --          & - \\  
 Adjusted CO                                    & 250 -- 275    & f$_{\rm sed}$ = 4, 6, 8 &   --                   &       --          & CO: -7.5 -- -5       \\  
 Adjusted CO$_{2}$                              & 250 -- 275    & f$_{\rm sed}$ = 4, 6, 8 &   --                    &       --          & CO$_{2}$: -10 -- -8.5\\  
 Adjusted PH$_{3}$                              & 250 -- 275    & f$_{\rm sed}$ = 4, 6, 8 &    --                     &       --          & PH$_{3}$: -10 -- -8.5\\ 
 Adjusted CO$_{2}$, PH$_{3}$, fixed CO           & 250 -- 275    & f$_{\rm sed}$ = 4, 6    &  --                    &       --          & CO = -7, CO$_{2}$: -9 -- -7.5, PH$_{3}$: -9 -- -7.5  \\ \hline
 
 \multicolumn{6}{l}{\textbf{PICASO - Cloudy, Disequilibrium}}  \\ \hline
 Base Grid                                      & 250 -- 275    & f$_{\rm sed}$ = 4, 6, 8 & K$_{\rm zz}$ = 7  &       H          &        -              \\ 
 Adjusted CO                                    & 250 -- 275    & f$_{\rm sed}$ = 4, 6, 8& K$_{\rm zz}$ = 7    &       H          & CO: -7.5 -- -5       \\ 
 Adjusted CO$_{2}$                              & 250 -- 275    & f$_{\rm sed}$ = 4, 6, 8& K$_{\rm zz}$ = 7     &       H          & CO$_{2}$: -10 -- -8.5\\ 
 Adjusted PH$_{3}$                              & 250 -- 275    & f$_{\rm sed}$ = 4, 6, 8& K$_{\rm zz}$ = 7     &       H          & PH$_{3}$: -10 -- -8.5\\
 Adjusted CO$_{2}$, PH$_{3}$, fixed CO           & 250 -- 275    & f$_{\rm sed}$ = 4, 6, 8& K$_{\rm zz}$ = 7    &       H          & CO = -7, CO$_{2}$: -9 -- -7.5, PH$_{3}$: -9 -- -7.5  \\ \hline

 \multicolumn{6}{l}{\textbf{coolTLUSTY - Clear, Equilibrium}}  \\ \hline
 Base Grid                                      & 200 -- 450    & --            &  --                   & --                 &     --                  \\ \hline
  \multicolumn{6}{l}{\textbf{coolTLUSTY - Clear, Disequilibrium}}  \\ \hline
  Base Grid                                      & 200 -- 450    & --            & K$_{\rm zz}$ = 6     &       H          &     --                  \\ 
 Adjusted Lengthscales                          & 200 -- 450    & --           & K$_{\rm zz}$ = 6     & H, 0.1 x H, 0.01 x H &    --                   \\ \hline 
 \multicolumn{6}{l}{\textbf{coolTLUSTY - Cloudy, Equilibrium}}  \\ \hline
  Base Grid                                      & 200 -- 350    & CDAMPU = 2, 6        &  --                     &       --           &   --                    \\ \hline
  \multicolumn{6}{l}{\textbf{coolTLUSTY - Cloudy, Disequilibrium}}  \\ \hline
  Base Grid                                      & 200 -- 350    & CDAMPU = 2, 6       & K$_{\rm zz}$ = 6 &       H          &    --                   \\

\hline
\end{tabular}
\caption{Summary of model grids and their parameters used to fit the mean WISE 0855 JWST/NIRSpec time-series spectrum described in more detail within Section~\ref{grid-intro}. Each grid covers a specified range of effective temperatures. Some of the atmospheric model grids include clouds, convection-driven disequilibrium chemistry, and/or individually adjusted chemical abundances. The base grid of the PICASO, clear disequilibrium models was initially run with a broader effective temperature range than the other PICASO grids.\\\\$^{a}$This column lists the variable each grid uses to parameterize the presence of clouds within the model and the values the parameter was calculated at. If clouds are not applicable this is indicated with a dashed line. \\\\$^{b}$This column lists the variable used to parameterize the impact of convection-driven disequilibrium (DisEQ) chemistry. The eddy diffusion coefficient (K$_{\rm zz}$) is listed where it applies along with the assumed value. If not applicable a dashed this is indicated with a dashed line.\\\\$^{c}$The lengthscale over which the eddy diffusion coefficient was calculated. H is the atmospheric scale height used in the approximation.\\\\$^{d}$ For each relevant molecule the range of adjusted log volume mixing ratios is listed.} 
\label{tbl:models}
\end{table*}

\subsection{Mean Spectrum Fitting}
\begin{figure*}
    \centering
    \includegraphics[width=7in]{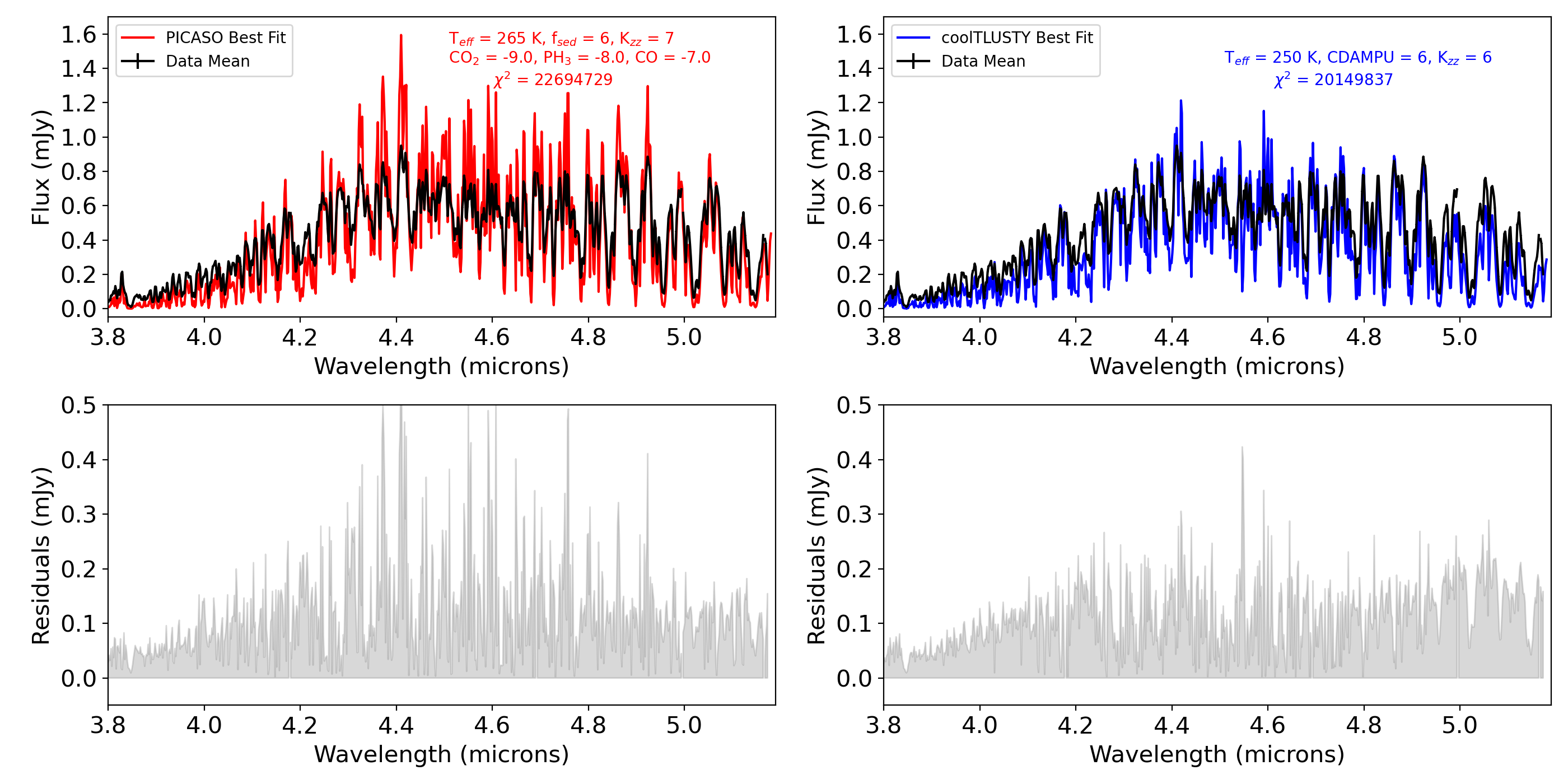}
    \caption{\textbf{Top Panels:} Best fit models from the PICASO (red) and coolTLUSTY (blue) grids to the mean JWST/NIRSpec time-series spectrum of WISE 0855 in black. \textbf{Bottom Panels:} Residual flux between the best fit model and the mean spectrum plotted in gray.}
    \label{fig:best fit mean spectra}
\end{figure*}

Understanding ultracool brown dwarf atmospheres with 1-dimensional forward models is challenging and the atmospheric fitting in this work is a demonstration of how well current models can explain our data set. For each grid in Table~\ref{tbl:models}, we complete a $\chi^{2}$ test assuming an object radius of 1 R$_{\rm jup}$ and distance of 2.28 pc \citep{Kirkpatrick_2021}.  The best fit model from each grid to the mean WISE 0855 spectrum is summarized in Table~\ref{tbl:best fit}. Model grids including water clouds have a lower $\chi^{2}$ value compared to the cloudless counterpart in all cases. Cloudy coolTLUSTY models have lower $\chi^{2}$ values compared to cloudy PICASO models. This difference is primarily due to the water cloud extent and particle distributions assumed in the PICASO and coolTLUSTY models which will be covered in the discussion section. Among the PICASO grids, the best fit models include adjusted abundances of \ce{CO2} and \ce{PH3} even when disequilibrium chemistry is already accounted for. The disequilibrium models tend to underestimate the abundances of \ce{CO2} and \ce{PH3} need the fit the mean spectrum. When either \ce{CO2} or \ce{PH3} are adjusted alone, the inferred abundance tends to be much higher to overcompensate for the lack of the other molecule. The best fit PICASO and coolTLUSTY forward model for the mean WISE 0855 spectrum are shown in Figure~\ref{fig:best fit mean spectra}. Adjusting the abundances of CO, CO$_{2}$, and PH$_{3}$ improves the raw $\chi^{2}$ values, but the addition of cloud opacity provides the greatest improvement in fit to the overall spectral energy distribution. Future work is needed to fine-tune models specifically to WISE 0855's atmosphere.

\begin{table*}

\tiny
\centering
\begin{tabular}{l|cllcc|c|c}
 \multicolumn{8}{c}{\textbf{Best Fit Parameters}}           \\ 
 \textbf{Grid Description }                    & T$_{\rm eff}$ & Cloud            &  DisEQ               &   Mixing         & Adjusted                            & $\chi^{2}$   &  $\chi_{\nu}^{2}$     \\ 
                                               &   (K)         & Parameter        &Parameter             &  Lengthscale      & log(VMR)                            &              &                 \\\hline
 \multicolumn{8}{l}{\textbf{PICASO - Clear, Equilibrium}}  \\ \hline                                                               
Base Grid                                      & 250           & --               &  --                   &       --          & -                                         & 47788349 & 37510 \\  
Adjusted CO                                    & 250           & --               &  --                   &       --          & CO: -7                                    & 29324727 & 23036 \\ 
Adjusted CO$_{2}$                              & 250           & --               &  --                   &       --          & CO$_{2}$ = -9.0                           & 46402232 & 36451 \\ 
Adjusted PH$_{3}$                              & 250           & --               &  --                   &       --          & PH$_{3}$ = -8.5                           & 45701006 & 35900 \\  
Adjusted CO$_{2}$, PH$_{3}$, fixed CO           & 260           & --               &  --                   &       --          & CO$_{2}$ = -9.0, PH$_{3}$ = -7.5, CO = -7.0& 25105979 & 19753 \\ \hline
\multicolumn{8}{l}{\textbf{PICASO - Clear, Disequilibrium}}  \\ \hline   
Base Grid                                      & 260           & --               &  K$_{\rm zz}$ = 7     &       H          & --                                         & 35359273 & 27820        \\ 
Adjusted CO                                    & 250           & --               &  K$_{\rm zz}$ = 7     &       H          & CO = -7                                    & 29238387 & 23004      \\ 
Adjusted CO$_{2}$                              & 260           & --               &  K$_{\rm zz}$ = 7     &       H          & CO$_{2}$ = -8.5                            & 32776934 & 25788  \\ 
Adjusted PH$_{3}$                              & 260           & --               &  K$_{\rm zz}$ = 7     &       H          & PH$_{3}$ = -8.5                            & 32743436 & 25762 \\ 
Adjusted CO$_{2}$, PH$_{3}$, fixed CO           & 260           & --               &  K$_{\rm zz}$ = 7     &       H          & CO$_{2}$ = -9.0, PH$_{3}$ = -7.5, CO = -7.0 & 26802762 & 21121 \\ \hline 

\multicolumn{8}{l}{\textbf{PICASO - Cloudy, Equilibrium}}  \\ \hline   
Base Grid                                      & 265           & f$_{\rm sed}$ = 6&  --                   &       --         & --                                         & 33921144 & 26647 \\  
Adjusted CO                                    & 270           & f$_{\rm sed}$ = 4&  --                   &       --         & CO = -7.0                                  & 24477153 & 19243  \\  
Adjusted CO$_{2}$                              & 265           & f$_{\rm sed}$ = 6&  --                   &       --         & CO$_{2}$ = -10.0                           & 34027082 & 26751 \\  
Adjusted PH$_{3}$                              & 265           & f$_{\rm sed}$ = 6&  --                   &       --         & PH$_{3}$ = -10.0                           & 33932924 & 26677 \\ 
Adjusted CO$_{2}$, PH$_{3}$, fixed CO           & 270           & f$_{\rm sed}$ = 4&  --                   &       --         & CO$_{2}$ = -9.0, PH$_{3}$ = -8.5, CO = -7.0& 24471379  & 19269 \\ \hline

\multicolumn{8}{l}{\textbf{PICASO - Cloudy, Disequilibrium}}  \\ \hline
Base Grid                                      & 270           & f$_{\rm sed}$ = 4& K$_{\rm zz}$ = 7      &       H          &    --                                      & 31740960 &  24954 \\ 
Adjusted CO                                    & 265           & f$_{\rm sed}$ = 8& K$_{\rm zz}$ = 7      &       H          & CO =  -7.0                                 & 23318547 & 18347  \\ 
Adjusted CO$_{2}$                              & 270           & f$_{\rm sed}$ = 4& K$_{\rm zz}$ = 7      &       H          & CO$_{2}$ = -8.5                            & 30632989 & 24101  \\ 
Adjusted PH$_{3}$                              & 270           & f$_{\rm sed}$ = 6& K$_{\rm zz}$ = 7      &       H          & PH$_{3}$ = -7.5                            & 26893022 & 21159    \\
Adjusted CO$_{2}$, PH$_{3}$, fixed CO           & 265           & f$_{\rm sed}$ = 6& K$_{\rm zz}$ = 7      &       H          & CO$_{2}$ = -9.0, PH$_{3}$ = -8.0, CO = -7.0& 22694729 & 17884   \\ \hline 

\multicolumn{8}{l}{\textbf{coolTLUSTY - Clear, Equilibrium}}  \\ \hline
Base Grid                                      & 250           & --               &  --                   &       --         &      --                                    & 56134390  & 44061   \\ \hline
 \multicolumn{8}{l}{\textbf{coolTLUSTY - Clear, Disequilibrium}}  \\ \hline
Base Grid                                      & 250           & --               & K$_{\rm zz}$ = 6      &       H          &      --                                    & 32199111  & 25294       \\ 
Adjusted Lengthscales                          & 250           & --               & K$_{\rm zz}$ = 6      & 0.01 x H          &        --                                  & 29275549 &   22997        \\ \hline 
\multicolumn{8}{l}{\textbf{coolTLUSTY - Cloudy, Equilibrium}}  \\ \hline
 Base Grid                                     & 300           & CDAMPU = 2       &  --                   &     --           &      --                                    & 22116156 &  17373    \\ \hline
 \multicolumn{8}{l}{\textbf{coolTLUSTY - Cloudy, Disequilibrium}}  \\ \hline
 Base Grid                                     & 250           & CDAMPU = 6       & K$_{\rm zz}$ = 6      &       H          &        --                                  &20149837 &   15829

\end{tabular}
\caption{The best fit model and parameters from each model grid in Table~\ref{tbl:models}. The chi squared ($\chi^{2}$) value for each model and the reduced chi squared values ($\chi_{\nu}^{2}$) are shown.}
\label{tbl:best fit}
\end{table*}

\subsection{Time-Series Spectra Fitting}
\label{grid-intro}
The cloudy disequilibrium atmospheric models from the PICASO and coolTLUSTY grids are used to interpret the modulations in the WISE 0855 time-series spectra. To avoid a large number of models run, the cloudy disequilibrium PICASO grid where \ce{PH3} and \ce{CO2} are adjusted with a fixed \ce{CO} are stitched at 4.45 $\mu$m with the adjusted CO grid. This allows the abundances of  \ce{PH3}, \ce{CO2}, and \ce{CO} to be changed simultaneously during spectral fitting. For each spectrum along the time-series a best fit model is found by interpolating over the model parameters and minimizing $\chi^{2}$ with the scipy \citep{2020SciPy-NMeth} function \texttt{minimize}. To simulate measurement noise a random draw is done treating the best fit spectrum as the mean and the error of the data as the standard deviation of the distribution. 

For the PICASO models, we tested four different cases to match the time-series variations: 

\begin{itemize}
    \item \textbf{1)} changing:  T$_{\rm eff}$, CO, CO$_{2}$, and PH$_{3}$, fixed: f$_{\rm sed}$

    \item \textbf{2)} changing  T$_{\rm eff}$ and fixed f$_{\rm sed}$

    \item \textbf{3)} changing f$_{\rm sed}$ and fixed T$_{\rm eff}$

    \item \textbf{4)} changing both f$_{\rm sed}$ and T$_{\rm eff}$
    
\end{itemize} 

For the coolTLUSTY models, three different cases were tested: 
\begin{itemize}
    \item \textbf{1)} changing  T$_{\rm eff}$ and fixed CDAMPU

    \item \textbf{2)} changing f$_{\rm sed}$ and fixed CDAMPU

    \item \textbf{3)} changing CDAMPU and T$_{\rm eff}$
\end{itemize}

\begin{figure*}
    \centering
    \includegraphics[width=6in]{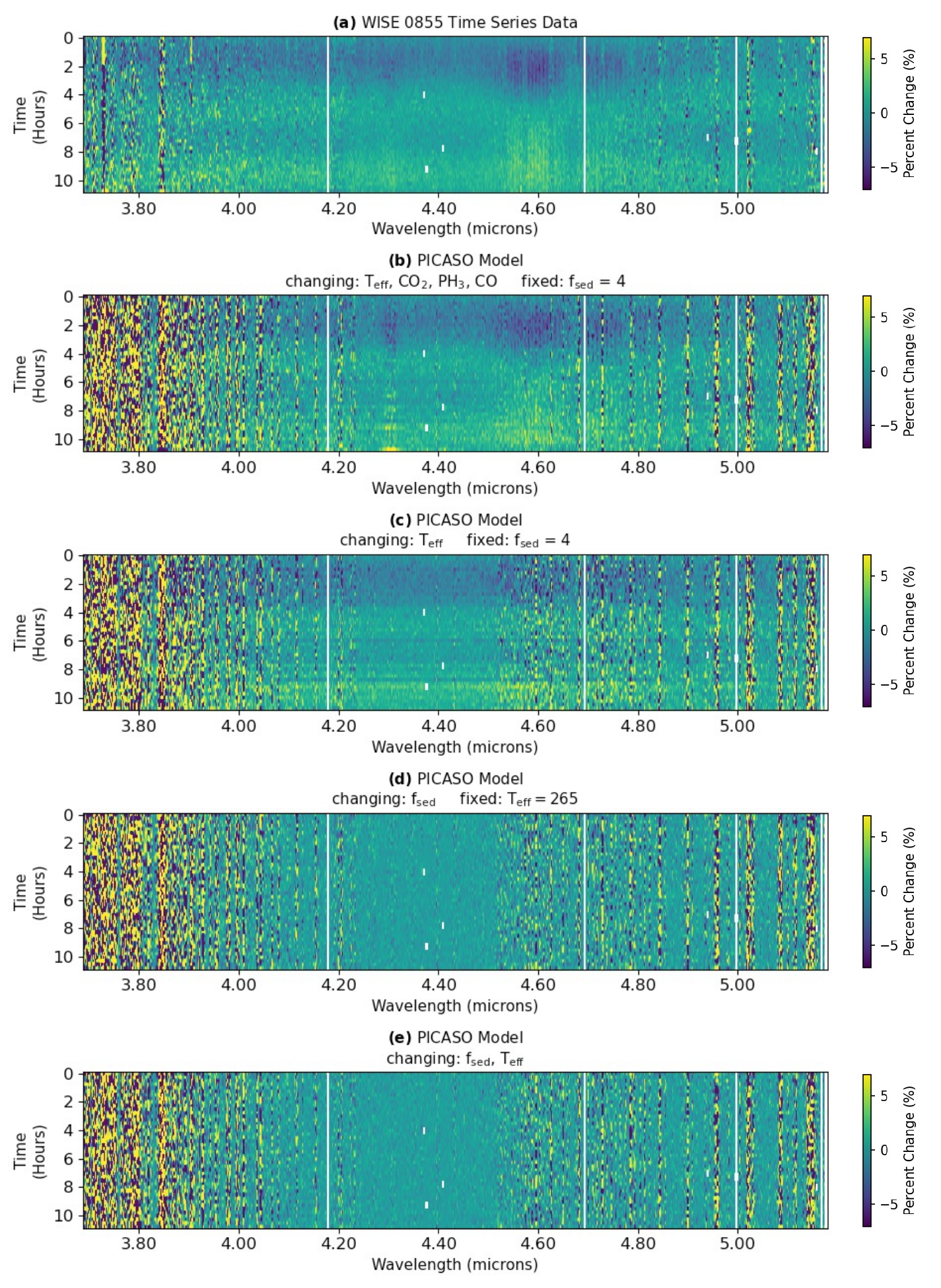}
    \caption{Percent change maps (blue-yellow) of the spectroscopic time-series observations \textbf{(a)} and PICASO models with various parameters changed. \textbf{(b)} Percent change map of the best fit PICASO model. The cloudiness parameter f$_{\rm sed}$ is fixed. Effective temperature and abundances of disequilibrium molecules (\ce{CO2, PH3, CO}) are changing. \textbf{(c)} Best fit percent change map changing only effective temperature with f$_{\rm sed}$ fixed. \textbf{(d)} Best fit percent change map allowing f$_{\rm sed}$ to change with fixed effective temperature. When f$_{\rm sed}$ is allowed to change alone it stays fixed. \textbf{(e)} Best fit percent change map allowing f$_{\rm sed}$ and effective temperature to change.}
    \label{fig:best fit percent change PICASO}
\end{figure*}

\begin{figure*}
    \centering
    \includegraphics[width=6.in]{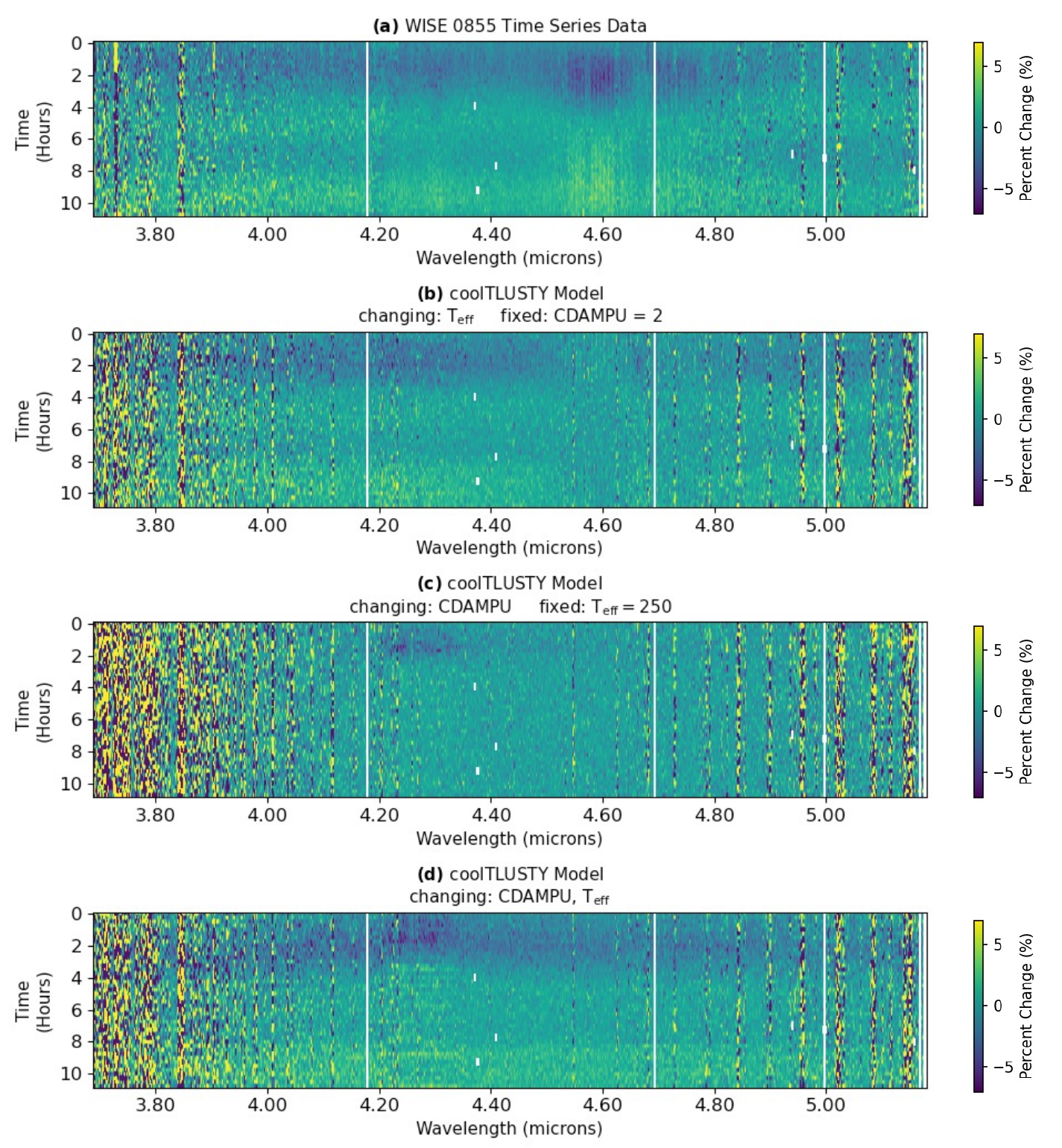}
    \caption{Percent change maps (blue-yellow) of the spectroscopic time-series observations \textbf{(a)} and coolTLUSTY models with various parameters changed. \textbf{(b)} Percent change map of the coolTLUSTY model where the cloudiness parameter CDAMPU is fixed and  effective temperature changes. \textbf{(c)} Best fit percent change map changing CDAMPU with effective temperature fixed. \textbf{(d)} The best fit percent change map allowing CDAMPU and effective temperature to change.}
    \label{fig:best fit percent change coolTLUSTY}
\end{figure*}

The percent change map of the best fit time-series spectra for each test case is shown in Figure~\ref{fig:best fit percent change PICASO} and the best fit parameters over time are plotted in Figure~\ref{fig:best fit percent change coolTLUSTY}. Qualitatively, the cloudy PICASO models with adjusted disequilibrium abundances capture the time variations within the CO and PH$_{3}$ disequilibrium features along with smaller amplitude changes that occur over the entire wavelength range of the data. Effective temperature is not a property that changes over time but is a proxy for the apparent heat visible across a rotating heterogeneous surface. The apparent effective temperature difference is 0.6 K over our time-series. The changes in apparent effective temperature mirrors the temporal behavior of the white light curve and the amplitude of the first eigenvector of the time-series. The best fit abundance of CO has a dynamic range of 0.05 dex covering log VMR of -6.88 to -6.83. It is important to note that when effective temperature decreases, CH$_{4}$ is preferred in the CO $\leftrightarrow$ \ce{CH4} reaction, whereas the opposite behavior is observed in the best fit parameters. The \ce{CO} abundance curve has a similar shape as the amplitude of the second eigenvector of the time-series. This suggests that the behavior of \ce{CO} originates from a different location on the surface than the temperature changes. When \ce{PH3} and \ce{CO2} are allowed to vary together only \ce{PH3} displays variations in log VMR between -8.38 and -8.18. The best fit abundance of \ce{CO2} appears fixed at log VMR~=~-9 because the full dynamic range of the data is not captured by the stitched model grid used.

Adjusting both effective temperature and cloud thickness (CDAMPU) for the coolTLUSTY models is required to recreate most of the relative changes observed in the time-series. The dynamic range of the best fit model effective temperatures is 1.3 K and the cloudiness parameter changes by 0.6. The cloudiness parameter changing from 4.0 to 4.6 approximately corresponds to the cloud deck thickness shrinking by half in altitude and decreasing in opacity. The PICASO and coolTLUSTY models suggest different average effective temperatures(264 K vs. 256
K), but show a similar trend over time that mirrors the white light curve. The cloud base for the coolTLUSTY models typically occurs at lower pressures than the PICASO models, because the condensation curve crossing pressure is determined by the P-T profile of the cloudless version of the models \citep{2023ApJ...950....8L}. This reinforces the work done in \cite{2022ApJ...927..184M} that emphasizes cloud parameterization is important for adequately modeling the changes in the time-series of WISE 0855 and other similar temperature Y-dwarfs.

\begin{figure*}
    \centering
    \includegraphics[width=6.5in]{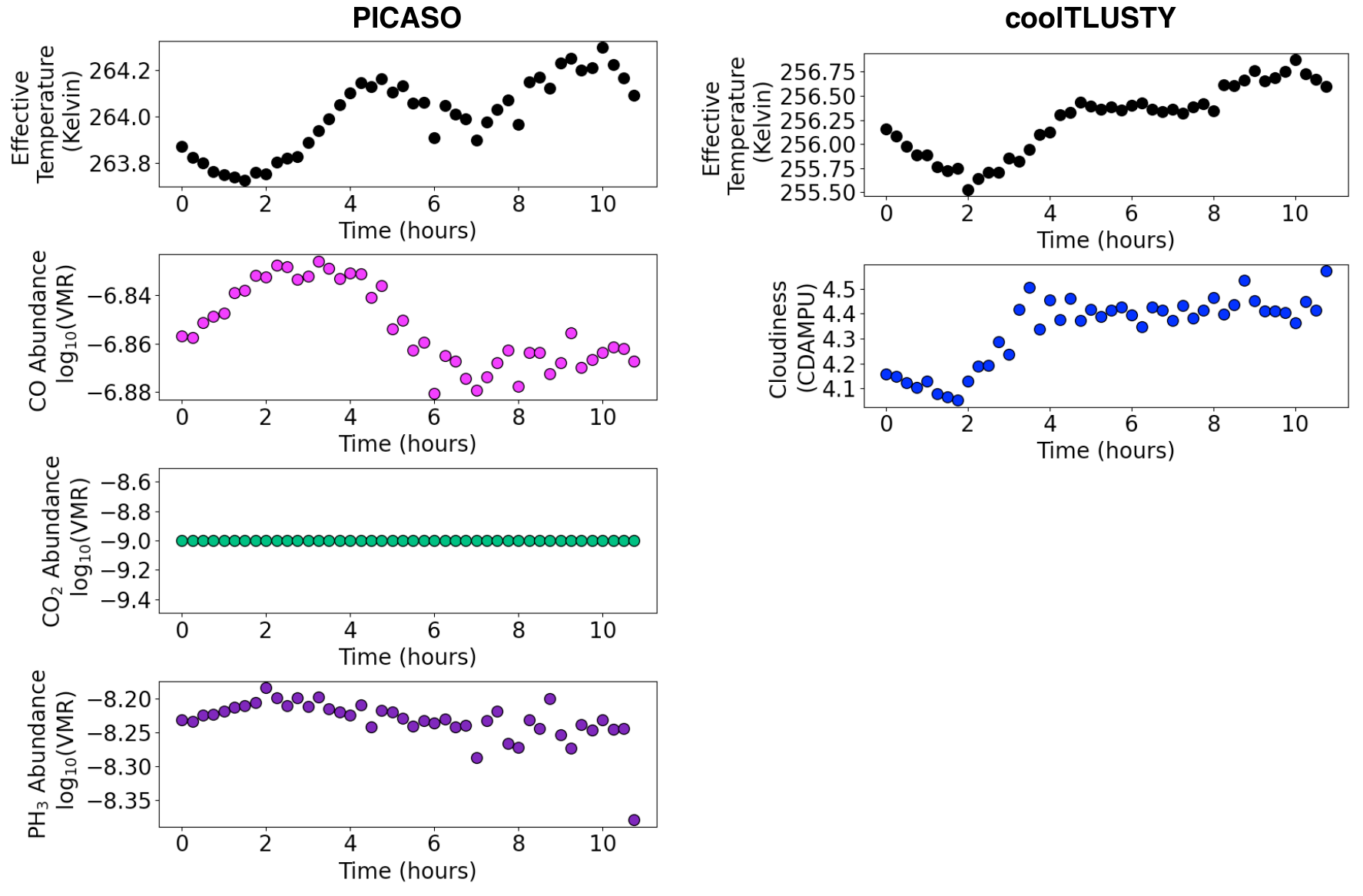}
    \caption{Best fit effective temperature and abundances of disequilibrium molecules over time for the best fit PICASO model from subplot b in Figure~\ref{fig:best fit percent change PICASO}. Best fit effective temperature and cloudiness parameter (CDAMPU) over time for the coolTLUSTY model in Figure~\ref{fig:best fit percent change coolTLUSTY}. The PICASO and coolTLUSTY models suggest different average effective temperatures (264 K vs. 256 K), but show a similar trend over time. The molecular abundances of disequilibrium molecules do not trend with effective temperature. In coolTLUSTY models CDAMPU corresponds to thinner clouds at higher values. }
    \label{fig:best fit parameter change PICASO}
\end{figure*}

\section{Discussion} \label{sec:Discussion}

\subsection{Disequilibrium Chemistry}
 Using the adjusted best fit abundances of \ce{CO}, \ce{CO2}, and \ce{PH3} in the PICASO models we can explore how 1-dimensional, vertical convection operates on WISE 0855 and understand if the behavior matches the assumptions of 1-dimensional atmospheric models. Similar to the methods in \cite{2020AJ....160...63M} we use the estimated abundances to determine the quench pressure for each molecule, then calculate an inferred eddy diffusion coefficient. The atmospheric structure model of the equilibrium version of the best fit mean PICASO model (T$_{\rm eff}$ = 265 K,  f$_{\rm sed}$ = 6, log(g) = 4) is assumed for all calculations. We find the pressure along the P-T profile where the thermochemical equilibrium abundance of a molecule matches the best fit adjusted abundance of the data and treat this as the quench pressure. The chemical timescale ($\tau_{\rm chem}$) is then calculated at the temperature and pressure of the quench point. We assume the mixing timescale ($\tau_{\rm mix}$) is equal to the chemical timescale and the relevant length scale is the pressure scale height (H) when calculating the inferred K$_{\rm zz}$ for each molecule. The following net chemical reactions for \ce{CO}, \ce{CO2}, and \ce{PH3} are assumed \citep{2006ApJ...648.1181V,2010Icar..209..602V}

 \begin{center}
    \ce{CO} + 3 \ce{H2} $\leftrightarrow$ \ce{CH4} + \ce{H2O} \\
    \ce{CO2} + \ce{H2} $\leftrightarrow$ \ce{CO} + \ce{H2O}   \\
    ~\ce{4PH3} + 6 \ce{H2O} $\leftrightarrow$ \ce{P4O6} + 12 \ce{H2}
\end{center}

\begin{figure*}
    \centering
    \includegraphics[width=7in]{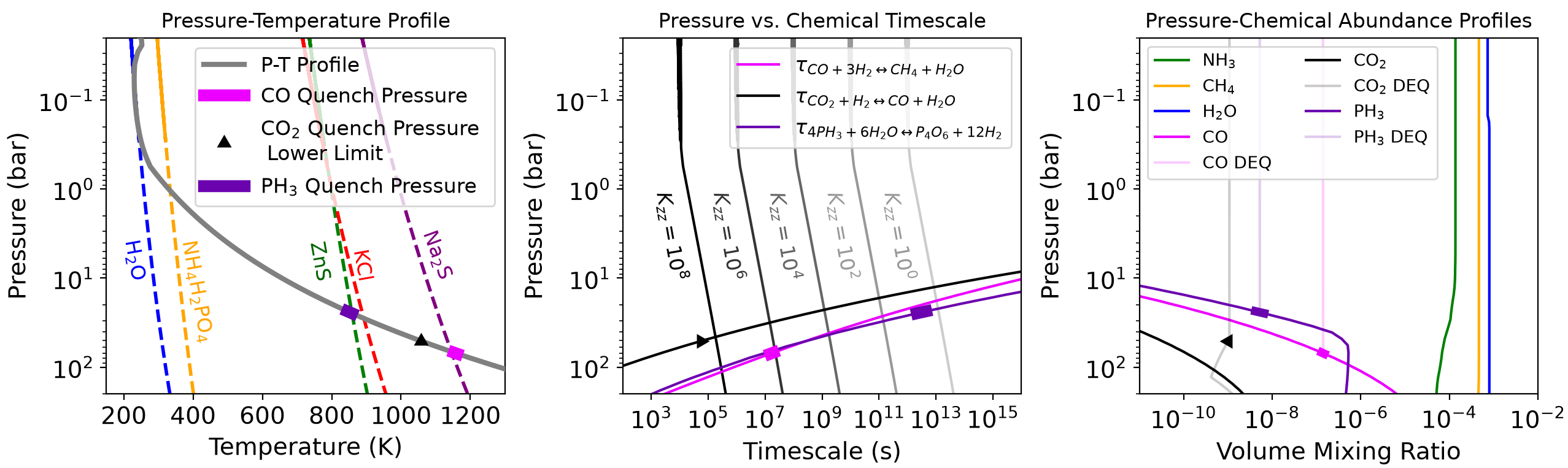}
    \caption{\textbf{Left:} Pressure-temperature profile of a 260 K, f$_{\rm sed}$ = 4 atmospheric model from PICASO (gray) with the condensation curves of several theoretical clouds plotted in colored dashed lines. The dynamic range of the estimated quench pressures for \ce{CO and PH3} are plotted in pink and purple respectively. The abundance of \ce{CO2} was not constrained therefore the lower limit on the quench pressure is plotted as a black triangle. \textbf{Center}: Chemical destruction timescales of \ce{CO,CO2 and PH3} plotted against pressure. Thick portions of the chemical timescales cover the quench pressures from the left panel. Each of the disequilibrium molecules have different inferred K$_{\rm zz}$ value ranges. Lines of constant eddy diffusion coefficient (K$_{\rm zz}$) are plotted in shades of black. \textbf{Right:} Abundances of various molecules versus pressure within the atmospheric model. The disequilibrium version of the abundances profiles are in faded colors.}
    \label{fig:chemistry}
\end{figure*}

The chemical timescale of the \ce{CO} $\leftrightarrow$ \ce{CH4} reaction is parameterized using equation (14) from \cite{2014ApJ...797...41Z}. The abundance of \ce{CO2} is dependent on the local \ce{CO} abundance, therefore the disequilibrium profile of \ce{CO} is used to determine the quench point \citep{2025RNAAS...9..108W}. Equation (44) from \cite{2014ApJ...797...41Z} is used to calculate the chemical timescale of the  \ce{CO2} $\leftrightarrow$ \ce{CO} reaction. For the oxidation of PH$_3$ \citep[e.g.,][]{2020JGRE..12506526V,2024ApJ...976..231L}, the chemical timescale for \ce{PH3} $\leftrightarrow$ \ce{P4O6} reaction is calculated using equation (20) of \cite{2005ApJ...623.1221V}. The inferred quench points and eddy diffusion coefficients of all three disequilibrium molecules are shown in Figure~\ref{fig:chemistry}. All disequilibrium species have inferred quench points within the same pressure scale height between 23.5 and 70.7 bars. In the 1-dimensional and constant K$_{\rm zz}$ vs. pressure case, these molecules should be influenced by the same bulk motion or have the same implied K$_{\rm zz}$ value. \ce{CO} has an inferred K$_{\rm zz}$ $\sim$ 10$^{6}$ $\mathrm{cm^{2}/s}$ and the inferred K$_{\rm zz}$ of \ce{CO2} is  at minimum $\sim$ 10$^{8}$ $\mathrm{cm^{2}/s}$. The inferred K$_{\rm zz}$ of \ce{PH3} is $\sim$ 10$^{1}$ $\mathrm{cm^{2}/s}$ which implies very little mixing.

Based on the clustering of light curves and principal component analysis, the behavior of \ce{PH3} and \ce{CO} mirror each other and are likely modulated together, not independently by any of the theoretical clouds \ce{ZnS}, \ce{KCl}, or \ce{Na2S}. The quench pressure of \ce{PH3} resides slightly above the theoretical condensation curves of zinc sulfide (\ce{ZnS}), potassium chloride (\ce{KCl}), and sodium sulfide (\ce{Na2S}) (Figure~\ref{fig:chemistry}). The quench pressure of \ce{CO} is deeper than all three predicted cloud species. The next deepest possible cloud is Manganese Sulfide (\ce{MnS}) which condenses at temperatures colder than $\sim$1550 K \citep{2006ApJ...648.1181V}. If we assume a maximum cloud extent of 0.5 dex in pressure \citep{2001ApJ...556..872A}, only CO, not \ce{PH3} could be modulated by its presence. The only clouds likely to be the source of inhomogeneities or patches that allow modulations of both \ce{CO} and \ce{PH3} to appear in our data are water and ammonium dihydrogen phosphate (\ce{NH4H2PO4}). The discrepancy in the estimated eddy diffusion coefficient of \ce{PH3} could be due to not knowing the primary reaction or sink of phosphorus based molecules. 

 Unlike \ce{PH3}, the limiting reaction of \ce{CO2} is directly dependent on the local abundance of \ce{CO} \citep[][]{2010Icar..209..602V} and they should share similar estimated values of K$_{\rm zz}$. The adjusted abundance PICASO grid does not go below a log(VMR) of -9, therefore we can only place a lower limit on the inferred K$_{\rm zz}$ value of \ce{CO2}. This lower limit is however, nearly two orders of magnitude larger than the inferred K$_{\rm zz}$ value of \ce{CO}. The relative abundance of \ce{CO2} and \ce{CO} is heavily impacted by metallicity \citep{2002Icar..155..393L} and in future work this can be explored to understand if it resolves some of the estimated vertical mixing tension between \ce{CO, CO2, and also PH3}. For now, assuming a solar abundance PICASO structure model, we do not derive the same K$_{\rm zz}$ for all disequilibrium molecules and vertical mixing strength does not increase with quench pressure. The inferred quench pressures and per molecule eddy diffusion coefficients are documented in Table~\ref{tbl:dis-eq summary}.

\begin{table}
\centering
\begin{tabular}{ c | c c }
Disequilibrium & Quench Pressure        & Inferred log(K$_{\rm zz}$)   \\ 
  Molecule        & Range (bars)   & Range log$(\mathrm{cm^{2}/s})$\\ \hline
 \ce{PH3} & 23.5 -- 24.8 & 1.37 -- 1.64 \\
 \ce{CO2} & $<$ 51 & $>$ 8.5 \\   
 \ce{CO}  & 68.7 -- 70.7 & 6.07 -- 6.18 \\ 
\end{tabular}
\caption{Estimated quench pressures and eddy diffusion coefficients of all disequilibrium molecules in WISE 0855 assuming a 265 K PICASO structure model. These molecules probe different pressures, but all quench within the same pressure scale height. The inferred K$_{\rm zz}$ values of \ce{CO2} and \ce{CO} are not consistent. The estimated K$_{\rm zz}$ value for \ce{PH3} is extremely small compared to \ce{CO2} and \ce{CO}. }
\label{tbl:dis-eq summary}
\end{table}

\subsection{The Presence and Composition of Clouds on WISE 0855}

By comparing our spectra to atmospheric models we can further interpret the contributions from individual molecular features (Figures~\ref{fig:features1}--\ref{fig:features4}) and broadband changes that could arise from cloud or temperature modulations. While the best fitting PICASO models include atmospheric clouds, the cloudiness parameter (f$_{\rm sed}$) remains constant when fitting the model grid over the time-series spectra. The observed changes in disequilibrium abundances from \ce{CO} and \ce{PH3} are too large to be caused by variations in cloud thickness or temperature alone. The abundance of \ce{H2O} gas is essentially constant at pressures larger than 5 bars, meaning the \ce{H2O} absorption lines are not impacted by vertical convection or changes in energy along the P-T profile (Figure~\ref{fig:chemistry}). This results in the PICASO model grid being unable to recreate the small variations in the data within the \ce{H2O} lines. In the PICASO framework, the upper level clouds are inhomogeneous but static, allowing regions of relatively warm gas to be observed along with modulations distinct to deeper pressure levels. Longitudinal gradients in atmospheric mixing (parameterized by K$_{\rm zz}$) are unlikely to explain these modulations because convective mixing is predicted to scale with rotational velocity, which only changes as a function of latitude \citep{2010Icar..209..602V}. This has also been supported with spatially resolved spectroscopic observations of Jupiter with Juno \citep{2020JGRE..12506206G}.

\begin{figure*}
\centering    
\includegraphics[width=7in]{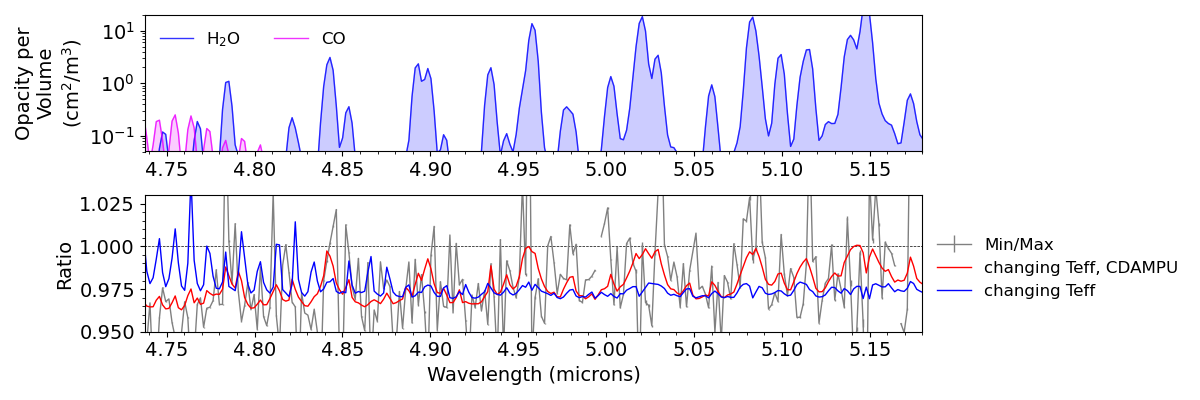}   \caption{\textbf{Top:} Opacity of significant molecules plotted vs. wavelength. Similar to Figures 4--7. Focusing on the region where water vapor is the dominant molecular absorber. \textbf{Middle:} Minimum brightness spectrum divided by the maximum brightness spectrum and binned by 5 pixels (gray). The ratio of the best fitting coolTLUSTY models where only effective temperature is allowed to change for the min and max spectra (Blue). The red line represents the same ratio, but effective temperature and cloudiness parameter are allowed to change. \textbf{Bottom:} The same plot as the middle panel, but at the native resolution of the NIRSpec data. The error bars of the ratio are plotted, but relatively small. Water clouds tend to mute variations within wavelengths where water vapor absorption is likely to be strong.}
\label{fig:min max water}
\end{figure*}

Both effective temperature and cloudiness are required to change in the best fitting coolTLUSTY model to recreate the broader wavelength variations in the time-series spectra, however these changes do not adequately fit the behavior of \ce{CO} and \ce{PH3}. In contrast to the PICASO models, the coolTLUSTY models are a better match to the variations in \ce{H2O} gas (Figure~\ref{fig:min max water}). The abundance of \ce{H2O} gas is relatively constant versus pressure except where \ce{H2O} clouds begin to condense out of the atmosphere. This indicates water clouds within WISE 0855's atmosphere reside at a lower altitude than what is assumed by the PICASO models. In the coolTLUSTY framework, WISE 0855's atmosphere is inhomogeneous due to water clouds of varying thickness. Here we conclude that WISE 0855's upper level clouds are likely composed of \ce{H2O} alone and regions of thinner clouds expose deeper pressure levels with enhanced abundances of \ce{CO} and \ce{PH3}.

Other clouds besides \ce{H2O} like \ce{ZnS}, \ce{KCl}, and \ce{Na2S} are predicted to exist deeper within the atmosphere of WISE 0855 \citep{2012ApJ...756..172M} and we discuss if our time-series spectra reveal evidence of their existence. As mentioned in the previous section, the quench pressures of \ce{CO}, \ce{CO2}, and \ce{PH3} each cross the condensation curves of \ce{ZnS}, \ce{KCl}, and \ce{Na2S} respectively. \ce{CO} and \ce{CO2} will not have the same inferred K$_{zz}$ value if there is a cloud layer between the two quench points. This suggests that the patches or holes in the upper level clouds either extend through the lower level sulfide and potassium clouds or those clouds are not present. The chemical timescale of \ce{PH3} is uncertain, but it is expected to be the dominant phosphorous-bearing molecule and could condense into \ce{NH4H2PO4} \citep[e.g.,][]{1994Icar..110..117F, 2006ApJ...648.1181V,2020JGRE..12506526V}. Given that \ce{PH3} and \ce{CO} share the same modulation behavior from principal component analysis in this data set we suggest the variations in disequilibrium molecules derives from high pressure temperature changes rather than deep patchy clouds. The retrieval work from  \cite{2025A&A...695A.224K} infers a potential cloud with a base at $\sim$ 10 bars. This base deviates significantly from any of the theoretically expected clouds and was inferred in a retrieval framework that disfavors changes in the P-T profile. 

\subsection{Atmospheric Features and Orientation of WISE 0855}
The rotational modulations observed in WISE 0855, other brown dwarfs, and Jupiter arise from inhomogeneities caused by vortices, zonal jets, or spots \citep{Showman_2019,2019AJ....157...89G}. The inclination of WISE 0855 is unknown, but we can infer the orientation and probable surface features based on our observations and theoretical expectations of Jupiter-like bodies. From our analysis there are at least two major contributions of variability: High altitude temperature or cloud variations and deep pressure temperature variations at or below the quench pressure of \ce{CO}. The $\sim$ 1 Kelvin variations in WISE 0855 are consistent with the temperature perturbations predicted in a 10 hour period simulation from \citep{2013ApJ...776...85S} which occurs across the entire visible hemisphere a brown dwarf.

\begin{figure}
\centering    
\includegraphics[width=3.5in]{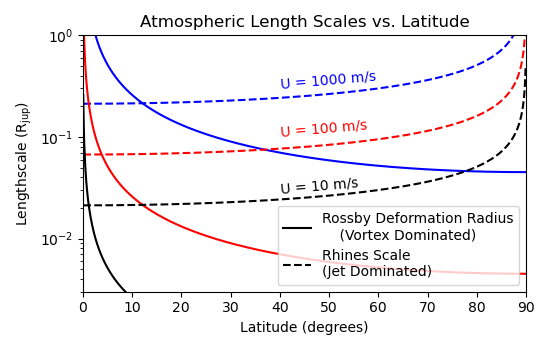}   \caption{Rossby deformation radius vs. Rhines scales as a function of latitude. We assume a rotation period of 11 hours and three different wind speeds (1000, 100, and 10 m/s). When the Rhines scale is larger than the Rossby deformation radius, zonal jets are expected to be the dominant surface feature. Vortices will dominate when the Rossby deformation radius is larger than the Rhines scale. Pole-on latitudes will lead to vortex dominated variations and edge-on latitudes will lead to jet dominated variations.}
\label{fig:lengthscales}
\end{figure}

WISE 0855 is unlikely to be in a pole-on orientation due to the measured large amplitude modulations from disequilibrium molecules \citep{2017ApJ...842...78V,2020AJ....160...38V,2021MNRAS.502.2198T, 2023ApJ...954L...6S}. These time variable spectroscopic features likely originate from zonal jets at equatorial to mid latitudes. Stable, large scale vortices only exist near polar regions and zonal jets are expected to dominate Y-dwarfs based on simulations \citep{2023MNRAS.525..150H}. Following \cite{2023MNRAS.525..150H}, we calculate the atmospheric length scales of both jets and vortices assuming three different horizontal wind speeds (10, 100, and 1000 m/s) (Figure~\ref{fig:lengthscales}). The characteristic atmospheric length scale of a vortex is only larger than a jet at latitudes of 78$^{\circ}$ and above considering the fastest wind speeds. Polar regions could potentially contribute to longer term variability rather than short term timescales of tens of hours \citep{2024ApJ...975L..32F}. There is currently no evidence of long term variations in the period based on the available Spitzer and JWST data. Future multi-epoch observations of WISE 0855 with JWST are required to determine if there is any polar structure or contribution to the observed variability. The contributions of \ce{CO} and \ce{PH3} occur on shorter timescales and are relatively sinusoidal. Given the relevant atmospheric length scales, we attribute these features to temperature variations along zonal jets. Longer baseline, multi-epoch observations with JWST are needed to determine if the deep temperature variations are the result of zonal jets offset by a combination of east-west zonal winds or a stagnant low latitude storm. 

\begin{figure*}
\centering    
\includegraphics[width=7in]{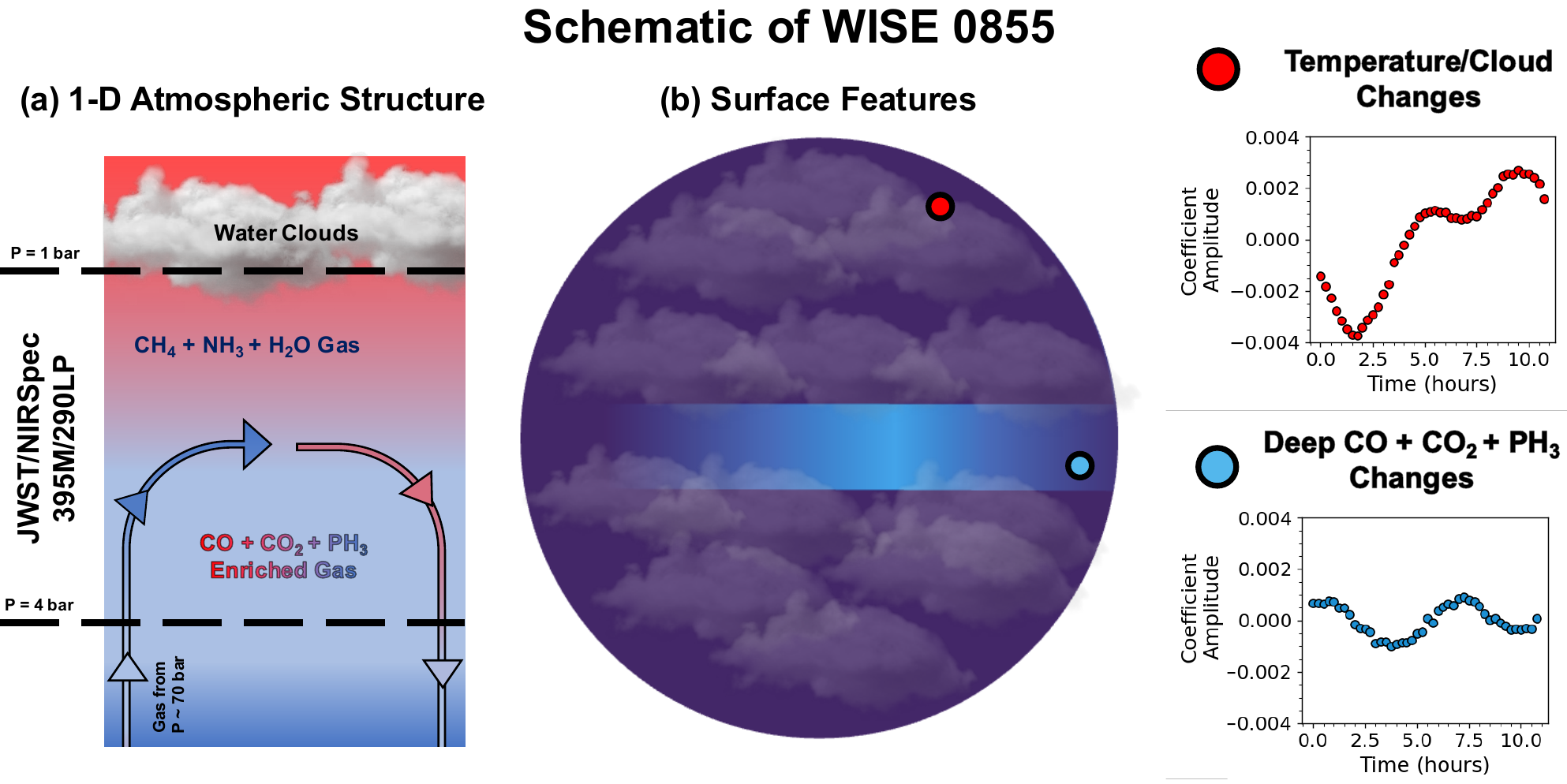}   \caption{Schematic of WISE 0855's atmosphere based on time-series observations and best fit models. \textbf{(a)} 1-dimensional representation of the atmosphere from top to bottom. The pressure sensitivity range of the JWST/NIRSpec mode is marked by the black dashed lines. Upper level clouds impact the wavelengths sensitive to the lowest pressures. Convection drives warm \ce{CO, CO2, and PH3} rich gas from high pressures into the photosphere. \textbf{(b)} Simplified 2-D representation of the surface features that could contribute to the observed variability. Purple regions are cloudy and modulate on $\sim$ 1 Kelvin scales. Blue regions have less cloud coverage and are sensitive to deeper pressures with stronger disequilibrium features. Both blue and purple regions can experience temperature modulations as described in \cite{2013ApJ...776...85S}. }
\label{fig:schematic}
\end{figure*}


\section{Summary}
In this paper we presented the first spectroscopic time-series observations of the Y-dwarf WISE 0855 using the BOTS mode of JWST/NIRSpec with the G395M/F290LP grating/filter setting. The 11-hour time-series spectra show variations with amplitudes ranging from 1 -- 10$\%$. In Section~\ref{sec:Features} we compared the time-series spectra to opacity curves of common molecules and find wavelengths with significant contributions from \ce{CH4, NH3, and H2O} display less variations relative to the white light curve. The largest amplitude variations occur within the \ce{CO} absorption bands. At wavelengths where the opacities of \ce{CH4, NH3, and H2O} are relatively low, modulations from other disequilibrium molecules such as \ce{CO2} and \ce{PH3} appear with strong variations. The changes in the apparent abundance of \ce{ PH3 and CO} are driven by deep temperature modulations within the convective region of the brown dwarf. Based on principal component analysis in Section~\ref{sec:Analysis}, 60.5\% of the variability can be represented by an eigenspectrum mimicking the behavior of apparent changes in effective temperature or cloud coverage. The next significant contribution (4.7\%) to the variability is an eigenspectrum which primarily modulates the absorption of \ce{CO and PH3}. In Section~\ref{sec:Analysis}, we tried to determine the rotation period of WISE 0855 using JWST and published Spitzer data. Within the JWST observation period, WISE 0855 shows significant periods between 5.4 and 5.6 hours. However, periods longer than the 11-hour observation window also have significant power. WISE 0855 has not shown any consistent periodicity over three different epochs with Spitzer and JWST. Longer baseline observations are likely needed to determine the true rotation period of this object. 

In Section~\ref{sec:Model Fitting}, we use various self-consistent atmospheric forward models for comparison with the JWST/NIRSpec data. When fitting the atmospheric models to the mean WISE 0855 spectrum water clouds are necessary in order to match the general spectral shape of WISE 0855's mean spectrum. The best fit to the mean spectrum was a cloudy, disequilibrium coolTLUSTY atmospheric model where T$_{\rm eff}$ = 250, CDAMPU = 6, log(K$_{zz}$) = 6, and the convective lengthscale is set equal to the pressure scale height. We find that changes in effective temperature and water cloud thickness are needed to explain the lack of variability within the water vapor absorption bands and broader wavelength modulations. Modulations in the abundances of \ce{CO and PH3} are needed on top of the variations in apparent temperature and cloud thickness to recreate the observed percent changes within these disequilibrium molecular bands. 

We can infer that the contributions of water cloud thickness and disequilibrium molecular gases must arise from different altitudes within the atmosphere based on PCA and model fitting analyses (Sections ~\ref{sec:Analysis}, \ref{sec:Model Fitting} and~\ref{sec:Discussion}). Based on the large amplitude \ce{CO} variations, theoretical expectations, and spatially resolved observations of Jupiter, we suggest that WISE 0855 is likely to be in an edge-on orientation. We estimated the strength of vertical mixing on a per-molecule basis in Section~\ref{sec:Discussion} and found that vertical mixing strength does not increase with higher pressure. Given that the local \ce{CO} abundance directly determines the abundance of \ce{CO2}, non-solar metallicities need to be explored in order to resolve the inferred K$_{\rm zz}$ tension between the molecules. The model estimated abundance of \ce{PH3} is small in value compared to theoretical expectations, which is consistent with the observed trend of \ce{PH3} depletion in brown dwarfs \citep{2024ApJ...973...60B,doi:10.1126/science.adu0401}.

Considering the relevant atmospheric lengthscales of storms and zonal jets in Section~\ref{sec:Discussion}, most of WISE 0855's latitudes are expected to be dominated by zonal jets rather than vortices. With the available data and theoretical models, we present a schematic of WISE 0855 in Figure~\ref{fig:schematic}. Using JWST/NIRSpec time-series observations, we have revealed that one of our coolest neighbors, WISE 0855 has complex atmospheric modulations driven primarily by varying disequilibrium molecules and water clouds.


\section*{Acknowledgments}
This work is based [in part] on observations made with the NASA/ESA/CSA James Webb Space Telescope. The data were obtained from the Mikulski Archive for Space Telescopes at the Space Telescope Science Institute, which is operated by the Association of Universities for Research in Astronomy, Inc., under NASA contract NAS 5-03127 for JWST. These observations are associated with program $\#$2327 and can be accessed via \dataset[doi:10.17909/j9rb-j449]{https://doi.org/10.17909/j9rb-j449}.

J.M.V. acknowledges from a Royal Society - Research Ireland University Research Fellowship (URF/1/221932, RF/ERE/221108) and the European Union through the Exo-PEA ERC project (grant number 101164652). Views and opinions expressed are however those of the author(s) only and do not necessarily reflect those of the European Union or the European Research Council Executive Agency. Neither the European Union nor the granting authority can be held responsible for them.

B.E.M. was supported by the Heising–Simons Foundation 51 Pegasi b Postdoctoral Fellowship and the University of Arizona's Presidential Postdoctoral Program for the duration of some of this work. B.E.M. is thankful for the invaluable emotional support from family, friends, and colleagues. B.E.M. dedicates this paper to her grandfather Alvin Turner Jr.

\bibliography{bibliography}
\bibliographystyle{aasjournal}



\end{document}